%% file: main.tex
\documentclass[10pt,conference]{IEEEtran}
\usepackage{booktabs}
\usepackage{seqsplit}
\usepackage{epsfig}
\usepackage[notextcomp]{stix}
\usepackage[font=small]{caption}
\usepackage{subcaption}
\usepackage{multirow, makecell}
\usepackage{latexsym}       % Make some extra symbols available
\usepackage{amsmath}        % Allows text in math mode among others
\usepackage{xspace}         % Smart spacing for macros
\usepackage{xcolor}         % Comment color
\usepackage{wasysym}        % Symbols
\usepackage{cite}           % Order citations by number
\usepackage{tabularx}       % Tables that span entire column
\usepackage{balance}        
\usepackage[utf8]{inputenc} % Unicode support
\usepackage{rotating}
\usepackage{textcomp}
\usepackage[notextcomp]{stix}
\usepackage[shortlabels]{enumitem}

\input{macros.tex}

\PassOptionsToPackage{hyphens}{url}
\usepackage{url}

\begin{document}
\title{Robust, privacy-preserving, transparent, and auditable on-device blocklisting}
\date{}

\author{
Kurt Thomas,
Sarah Meiklejohn,
Michael A. Specter,
Xiang Wang,\\\vspace{.2cm}
Xavier Llorà,
Stephan Somogyi,
David Kleidermacher\\
Google
}

\maketitle

\input{00abstract}
\input{01introduction}

\input{02background}
\input{03framework}
\input{04a-design}

\input{04b-rlwe-design}
\input{05a-evaluation}
\input{05b-deployment}
\input{06-limitations}
\input{07conclusion}
\input{acks}

%\balance
{
\footnotesize

\bibliographystyle{plain}
\bibliography{libraryBib}
}

%\balancecolumns

\input{appendix}

\end{document}

%% file: macros.tex
\renewcommand{\paragraph}[1]{\vspace{5pt}\noindent\textbf{#1.\xspace}}
\newcolumntype{R}{>{\raggedleft\arraybackslash}X}
\newcolumntype{L}{>{\raggedright\arraybackslash}X}

\newcommand{\eg}{e.g., }
\newcommand{\etc}{etc.}

\newcommand{\etal}{et al.\@\xspace}

\newcommand{\halfcircle}{$\circlelefthalfblack$\xspace}
\newcommand{\fullcircle}{\newmoon\xspace}
\newcommand{\emptycircle}{$\times$\xspace}

\newcommand{\randpick}{\xleftarrow{\scriptscriptstyle \$}\xspace}

\newcommand{\NN}{\mathbb{N}}
\newcommand{\FF}{\mathbb{F}}
\newcommand{\secp}{\ensuremath{\lambda}}
\newcommand{\usecp}{\ensuremath{1^\secp}}

\newcommand{\pk}{\mathsf{pk}}
\newcommand{\sk}{\mathsf{sk}}
\newcommand{\key}{\mathsf{key}}

\newcommand{\sig}{\mathsf{Sig}}
\newcommand{\keygen}{\mathsf{KeyGen}}
\newcommand{\sign}{\mathsf{Sign}}
\newcommand{\verify}{\mathsf{Verify}}

\newcommand{\se}{\mathsf{SE}}
\newcommand{\enc}{\mathsf{Enc}}
\newcommand{\dec}{\mathsf{Dec}}

\newcommand{\fhe}{\mathsf{FHE}}
\newcommand{\add}{\mathsf{Add}}
\newcommand{\mult}{\mathsf{Multiply}}
\newcommand{\absorb}{\mathsf{Absorb}}

\newcommand{\entry}{\mathsf{entry}}
\newcommand{\coms}{\mathsf{coms}}

\newcommand{\proveincl}{\mathsf{ProveIncl}}
\newcommand{\verincl}{\mathsf{VerIncl}}
\newcommand{\provecons}{\mathsf{ProveCons}}
\newcommand{\vercons}{\mathsf{VerCons}}
\newcommand{\chkpt}{\mathsf{chkpt}}

\newcommand{\idx}{\mathsf{idx}}
\newcommand{\obj}{\mathsf{obj}}

\newcommand{\entitystyle}[1]{{\ensuremath{\mathsf{#1}}}\xspace}

\def\req{\entitystyle{Req}}
\def\resp{\entitystyle{Resp}}

\def\objectshort{\entitystyle{o}}
\newcommand\harmfuldetect{\entitystyle{HarmfulDetect}}
\newcommand\harmfuleval{\entitystyle{HarmfulEval}}

\def\objone{\objectshort_1}
\def\objtwo{\objectshort_2}
\def\objset{\entitystyle{O}}

\def\clientone{\entitystyle{c_1}}
\def\clienttwo{\entitystyle{c_2}}

%% file: 00abstract.tex
\begin{abstract}
With the accelerated adoption of end-to-end encryption, there is an opportunity to re-architect security and anti-abuse primitives in a manner that preserves new privacy expectations. In this paper, we consider two novel protocols for on-device blocklisting that allow a client to determine whether an object (\eg URL, document, image, \etc) is harmful based on threat information possessed by a so-called remote enforcer in a way that is both privacy-preserving and trustworthy. Our protocols leverage a unique combination of private set intersection to promote privacy, cryptographic hashes to ensure resilience to false positives, cryptographic signatures to improve transparency, and Merkle inclusion proofs to ensure consistency and auditability. We benchmark our protocols—one that is time-efficient, and the other space-efficient—to demonstrate their practical use for applications such as email, messaging, storage, and other applications. We also highlight remaining challenges, such as privacy and censorship tensions that exist with logging or reporting. We consider our work to be a critical first step towards enabling complex, multi-stakeholder discussions on how best to provide on-device protections.
\end{abstract}

%% file: 01introduction.tex
\section{Introduction}
Blocklists---aggregate lists of threat information indicators pertaining to previously identified harmful content---are extensively used for security and anti-abuse as a basic protection primitive. Canonical examples include blocking inbound spam messages based on IP addresses~\cite{ramachandran2007filtering}, identifying inbound spam calls based on phone numbers~\cite{android-caller-id}, displaying warnings for known phishing URLs~\cite{akhawe2013alice}, preventing malware downloads based on file hashes~\cite{chrome-malware-warning}, or preventing the spread of harmful media~\cite{giftc-database, ncmec-database} via lists of perceptual hashes.

With the rise in popularity of end-to-end encryption for private messaging~\cite{messenger-encryption} and recent pushes to adopt similar technologies for email~\cite{MLS}, storage backups~\cite{apple_icloud_e2ee}, and even social media~\cite{facebook-encrypted}, there is a general need to build security and anti-abuse protections that adhere to these new privacy expectations. Current blocklist technologies rely on clients sharing identifiers with a server-side API (\eg a hash prefix of a URL), or servers having access to unencrypted user data (\eg files stored remotely). This server-centric model is anchored in two rationales: (1) sensitivity around making blocklist elements readily available to attackers, which would otherwise provide an efficient oracle for improving evasion; and (2) efficiency due to more powerful server-computing resources. During this paradigm shift, where increasingly powerful devices emerge with stricter privacy expectations, we argue there is an opportunity to develop new on-device blocklist primitives which are privacy-preserving and trustworthy. 

The push for on-device blocklisting has not escaped the attention of researchers. A variety of proposals now exist for privately querying URL blocklists~\cite{kogan2021private}, passwords exposed in data breaches~\cite{thomas2019protecting, li2019protocols}, or whether an image contains harmful media such as child sexual abuse~\cite{appleneuralhash, hua2021increasing, kulshrestha2021identifying, riazi2016sub}. Critically, none of these proposals address transparency around blocklist entries, a feature that prevents on-device blocklisting from silently turning into a tool for censorship or surveillance---a key concern of security and privacy researchers~\cite{abelson_bugs_2021}. Additionally, proposals take varying stances about exactly who is being protected---users or services---and the resilience of any blocklisting to false positives---particularly those induced by adversarial examples.

In this paper, we propose a set of design principles that any on-device protection should achieve in order to be considered \emph{robust, privacy-preserving, transparent, and auditable}. This includes assuring user data privacy, blocklist privacy, and the resulting verdict's privacy, even in the presence of collusion. Equally important, our proposed design ensures that clients receive proof of the origin of blocklist entries, proof of consistent enforcement across all devices and applications, and proof that an auditor can reconstruct the history of all elements—past and present—in any blocklist. The precision of on-device blocklisting also hinges on false positives being computationally impractical to produce by an attacker. To the best of our knowledge, all existing protocols fail to meet at least one of these principles.

To address this gap, we outline two candidate protocols that, if implemented correctly, achieve all of the aforementioned properties. We intentionally use off-the-shelf, proven cryptographic techniques in our proposal to facilitate near-term adoption. In doing so, we augment existing strategies for private set intersection to include multiple cryptographic proofs that achieve our transparency principles. These include signatures on the origin of the blocklist entries, Merkle inclusion proofs that ensure blocklist entries are consistent across every device, and exportable proofs that allow the user to publicly disclose and contest entries or a cryptographic collision. We provide benchmarks to demonstrate that these techniques are computationally practical for real-world deployment: achieving near real-time performance in one case, and space-efficient performance for the other. We conclude with a discussion of the key remaining challenges our community should address in order to make on-device blocklisting feasible, including the risks that exist with migrating security and anti-abuse protections onto devices.

To be clear, we take no position on whether or not the protocols discussed in this paper should be adopted in practice. These protocols are only a step towards exploring on-device protections, which can evolve over time and adapt to changing landscapes. Systems that do not maintain user privacy, leak or report the contents of messages without user consent, or remain unaccountable as to the contents of their blocklists put strain on user trust and safety. Our principles are consistent with Android's multiparty consent model~\cite{mayrhofer2021android} and Google’s AI Principles~\cite{google_ai_principles}. Complex, multidisciplinary discussions are still needed to drive a consensus towards how or whether on-device blocklisting is appropriate, particularly in the context of end-to-end encryption. Our design principles and overview of the current landscape of proposals and their limitations can help shape the discussion for how best to achieve on-device protections while preserving communication privacy.

%% file: 02background.tex
\section{Background \& Related Work}
Here we provide an overview of client-side and server-side protections, their limitations, and key technologies that researchers have proposed as solutions in this space. 

\subsection{Background} \label{sec:background}

\paragraph{Definitions of E2EE} There have been a number of works that attempt to define end-to-end encryption from both technical and policy perspectives. The earliest definitions consider ``end to end’’ largely as an argument for the design of networking protocols---that routers should largely ignore the content of packets and allow the application layer to take care of the rest~\cite{clark}. Unfortunately, this  focuses on device-to-server communication (e.g. TLS) vs user-to-user communication (e.g. RCS or Signal). Others have since provided modern definitions E2EE, including the U.S.’s Federal Trade Commission~\cite{FTC_ZOOM}, academic work such as Hale and Komlo~\cite{hale_komlo}, and IETF standards drafts by Knodel \etal~\cite{knodel} and Muffett~\cite{muffett}. 

In this paper, we adopt the definition of E2EE provided by Muffett, and aim to maintain these definitional principles~\cite{muffett}. Intuitively, Muffett’s definition ensures that all participants have equal access to plaintext, that no non-participants have access to message contents, and that the list of participants are available to all participants before a message is sent, unless a participant intentionally takes specific action to forward or retransmit a message.

\paragraph{Client-side vs. server-side protections} Client-side protections are presently deployed in scenarios where security, privacy, or other practical considerations preclude the possibility of middleboxes or traditional server-side solutions (e.g. anti-virus scanning and spam call blocking). 
These protections entail that the device engage in a remote lookup, or otherwise perform some on-device processing, in order to arrive at a verdict. Client-side protections also help to overcome split view or time-of-check versus time-of-use attacks. For example, malicious content on the web is often \emph{cloaked}, appearing benign to security crawlers, but delivering scams and malware when rendered on a user's device~\cite{wang_cloak_2011}.

Current server-side strategies for preventing spam, malware, and other abuse hinge on access to content and fail to directly translate to end-to-end encrypted settings. Server-based techniques---such as email spam filtering---leverage course-grained analyses of user content and often entail manual review to ensure accurate detection. Privacy expectations for users of end-to-end encryption are clearly different, particularly surrounding what content leaves a device, constraining similar detection regiments. This is critically relevant as researchers consider end-to-end-encryption for email~\cite{MLS} and online social networks~\cite{feldman_social_2012, barenghi_snake_2014, persona}. 

Without content, some service providers rely on limiting functionality within end-to-end encrypted environments to prevent abuse~\cite{pfefferkorn_content-oblivious_2021}. For example, some messaging applications have responded to misinformation by limiting the number of members that can participate in large-scale groups, by restricting users' ability to forward messages~\cite{hitchen_analysis_2019}, and by relying on metadata around user messaging habits~\cite{whatsapp-forwarding}. 

\paragraph{Challenges with existing client-side protections} Current client-side protections do not provide a mechanism to contest or externally prove that the system reached a particular decision. This makes them vulnerable to undebuggable failures, breaches of trust by the security provider, or arbitrary or inconsistent decisions on what constitutes abuse~\cite{abelson_bugs_2021, kud,kamara_outside_2022}. 

This legitimate need for transparency is also in tension with the need for secrecy surrounding anti-abuse systems. For example, the popular E2EE messaging app Signal---though otherwise entirely open source---relies on a closed-source system for spam prevention. The developers have highlighted the difficulty of maintaining a transparent anti-abuse system in the face of an active adversary: ``If we put this code on the Internet alongside everything else, spammers would just read it and adjust their tactics to gain an advantage in the cat-and-mouse game of keeping spam off the network''~\cite{signal}. 

Additionally, it is likely that future improvements to privacy-preserving systems will complicate the use of content-agnostic tactics. Work on metadata-private messaging, for example, has shown that messaging systems need not maintain information about users or their messaging habits to provide basic functionality~\cite{kwon_riffle_2016, lazar_alpenhorn_2016}.

\subsection{Related Work}\label{sec:related}

\paragraph{Private Set Membership} A common strategy for guaranteeing privacy has been to leverage various private set membership (PSM) primitives, first introduced by Chor \etal~\cite{chor1995private}. These techniques allow a client to query a database controlled by a server, while ensuring the client's query and all other database elements remain secret. The main limitation of PSM is that computation is linear to the size of the server's database when only a single server is involved~\cite{haitner2008linear}. In order to minimize overhead, previous protections that use PSM have opted to leak a hash prefix of metadata or of the content itself, allowing the server to partition its database and reduce the search space. Examples include leaking some bits of a username when privately querying the breach status of a username and password pair~\cite{thomas2019protecting, li2019protocols}, or a partial hash of content in order to support near-duplicate lookups of perceptual hashes~\cite{hua2021increasing}. Partial leakage of this nature is likely to be at odds with the privacy assumptions of end-to-end encrypted environments.

\paragraph{Homomorphic Encryption \& Perceptual Matching} In order to overcome the exact matching limitations of PSM, researchers have leveraged homomorphic encryption techniques, specifically for near-duplicate matching of perceptual hashes like PhotoDNA or PDQ~\cite{photodna, facebook-pdq}. Singh \etal~\cite{singh2019robust} considers a setting in which the user sends an encrypted image to the server, then the server leverages the additive homomorphic properties of the ciphertext to discover if there is a match. Kulshrestha and Mayer~\cite{kulshrestha2021identifying} build on Singh~\etal, expanding protections to users in a setting where the server learns nothing more than the output of the query.  Their protocol allows the user’s device to obliviously calculate a hamming distance between images, and hashes are retrieved by the client using a private information retrieval scheme. 

In an end-to-end encrypted setting, such perceptual hashing techniques do not satisfy a number of desirable properties, although these techniques have proven effective in server-side environments.\footnote{Further references to perceptual hashes in this paper will similarly be specific to the on-device context. As noted, server-side use cases for perceptual hashing are well documented and have been utilized effectively, especially when coupled with other techniques.} There have been quite a few works demonstrating that it is possible to invert hashes, generate collisions, and that minimal modification of images can result significantly different hash values~\cite{reverse-photodna, hao2021s, struppek2021learning, jain2021adversarial, wang2021prototype, dolhansky2020adversarial, bai2020targeted}. To the best of our knowledge, creating a perceptual hash system that does not suffer from these vulnerabilities---absent manual verification post-detection or other server-side secrecy protections---is an area of open research.

\paragraph{Multiparty Computation} Multiparty computation assumes the existence of multiple, non-colluding servers to improve on the two party computation model of PSM. For example, Kogan \etal explored efficient blocklisting of phishing URLs without any privacy leakage, though at the cost of assuming at least one trusted server~\cite{kogan2021private}. Likewise, Riazi \etal proposed a nearest-neighbor search that relied on two non-colluding servers~\cite{riazi2016sub}. We consider this to be an unproven trust assumption for end-to-end encrypted environments: were the non-collusion assumption violated, the protocols would reveal the client's query without any privacy protections.

\paragraph{Trusted Execution Environments} A number of works study the use of hardware-supported trusted execution environments, particularly in the context of network middleboxes, as an aid to protect against spam and malware~\cite{han2017sgx, goltzsche_endbox_2018,sherry_blindbox_2015}. Given the recent attacks on SGX~\cite{van2018foreshadow}, we consider this an unproven trust assumption compared to the expectations of end-to-end encrypted environments.

\paragraph{Threshold Encryption} Threshold encryption ensures that an encrypted object can only be decrypted after obtaining access to at least $n$ partial keys. Bhowmick~\etal leveraged threshold encryption to report every abuse incident to a server, but the data associated with each report would become accessible only after a predefined number of violations~\cite{bhowmick_apple_nodate}. Such a reporting threshold could also be accomplished client-side via a stateful counter and local storage of the content, but is potentially vulnerable to a malicious client tampering with the state.

\paragraph{Message Franking} The goal of message franking is to disincentivize abusive behavior in E2EE messaging by selectively puncturing the system's deniability property, allowing specific third parties (\eg the service provider) to cryptographically verify the authenticity of a particular message~\cite{tyagi2019asymmetric,alhaddad_hecate_2021}. Techniques have been developed for both the use case of user-to-user chat, as well as for traceback protocols which allow the platform owner to discover the originator of a message through various message forwarding steps~\cite{tyagi2019traceback}. We view message franking as both orthogonal (in that it does not block abuse directly) and complementary to blocklisting efforts.

%% file: 03framework.tex
\section{Threat Model and Design Principles}\label{sec:framework}

Before discussing our protocols for on-device blocklisting, we define our threat model and the design principles that guide our approach. We compare these constraints against a variety of existing privacy-preserving protocols for on-device blocklisting to situate our work in the broader design space.

\subsection{Abstract protocol}
\begin{figure}[t]
    \centering
    \includegraphics[width=\columnwidth]{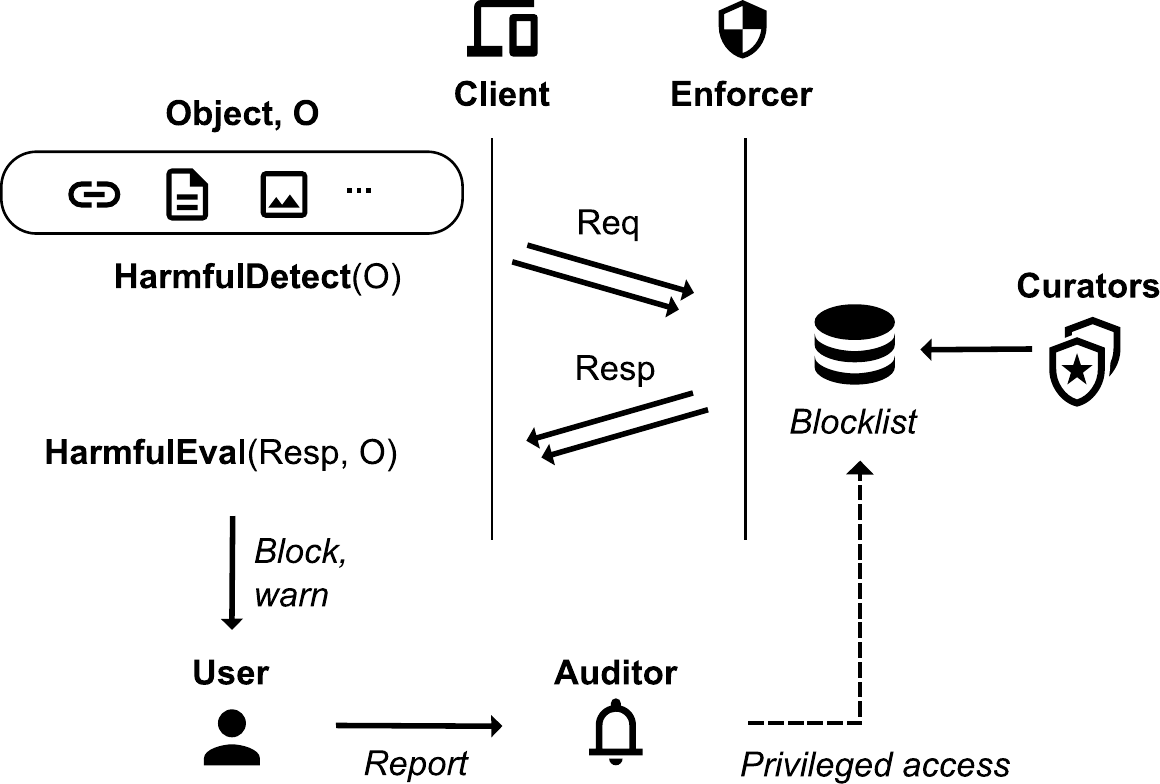}
    \vspace{1pt}
    \caption{Abstract protocol for performing on-device blocklisting. A client and enforcer exchange messages in an interactive protocol, $\harmfuldetect$, to determine whether or not an object $\objectshort$ held by the client is harmful. The enforcer has access to a blocklist, populated by threat intelligence from one or more curators. At the end of the interaction, during which it has received some response $\resp$ consisting of the messages sent by the enforcer, the client determines the blocklist status of the object by running $\harmfuleval(\objectshort, \resp)$. Positive verdicts result in blocking harmful content on the client, or otherwise warning the user. An auditor helps ensure correctness of the blocklist.\vspace{-15pt}}
    \label{fig:abstract_protocol}
\end{figure}

We outline an abstracted protocol for on-device blocklisting in Figure~\ref{fig:abstract_protocol}. 
The goal of on-device blocklisting is for a \emph{client} (\eg mobile app) to obtain a verdict via a secure channel from an \emph{enforcer} on whether or not an \emph{object} \objectshort (\eg URL, document, image, \etc) is harmful content.  We adopt the term harmful as it encompasses multiple potential threats, including scams, phishing, malware, child sexual abuse, violent extremism, misinformation, and other forms of abuse.\footnote{As we show, extending on-device blocklisting to a category of harmful content requires robust infrastructure, curated threat information indicators, and auditability. While in concept this technology is extensible, in practice, only a handful of content types today (\eg phishing, child sexual abuse material) meet these requirements.}

We define the interaction between the client and the enforcer as an interactive protocol $\harmfuldetect$, in which messages sent by the client are called \emph{request} messages (and denoted $\req$) and messages sent by the enforcer are called \emph{response} messages (and denoted $\resp$).  Enforcers determine the blocklist status of an object based on the consensus of one or more threat intelligence \emph{curators}.\footnote{An enforcer may also be a curator, but we differentiate these roles to clarify the responsibilities of a party in either context.} At the end of an interaction for an object \objectshort, a client runs $\harmfuleval(\objectshort, \resp)$. Upon a positive verdict, the client can opt for context-specific enforcement such as outbound protections (\eg blocking an object from being remotely distributed) or inbound protections (\eg prevent rendering or access to the object, or displaying a warning). For clarity, we refer to the operator of the client as the \emph{user}. This distinction is necessary for scenarios where a client enforces blocklist verdicts irrespective of the user opting-in to protection; as well as to differentiate reporting initiated by the client versus the user.

An \emph{auditor} provides an independent assessment of the correctness of verdicts. We envision two types of auditors: those that are \emph{privileged} with access to the set of blocklist elements as well as any intermediate signals or the original harmful objects that enforcers and curators use to arrive at verdicts, and \emph{unprivileged} auditors that may only receive a transcript of the user's interaction with the enforcer, provided directly by the user.

The main goal of the unprivileged auditor is to provide a means for any user to seek \emph{redress} via any entity they trust in order to hold the system to account, should the user encounter a malicious or erroneous addition to the database, or denial of service attack whereby the enforcer prevents clients from accessing the blocklist. Unprivileged auditors need only to have access to public parameters of the system to cryptographically verify the user's complaint. We envision a wide variety of potential unprivileged auditors, including judges, journalists, or the general public. Conversely, a privileged auditor would require a relationship with the enforcer; we avoid any initial recommendations on whom civil society or enforcers would advocate for such a role.

\subsection{Threat model}
We assume that the user, client, enforcer, any curators, and auditors operate in an adversarial fashion. Users and clients are honest-but-curious, meaning they honestly follow the protocol and obtain the latest, updated implementation, but attempt to learn as much as possible in order to evade detection or learn information about the contents of the blocklist. For example, a phishing campaign operator may masquerade as a client in order to learn the blocklist status of all of their phishing pages. However, users or clients that circumvent the blocklist check, for example by tampering with the client's execution, are explicitly outside the scope of our threat model. We defer a detailed discussion of such attacks until Section~\ref{sec:limitations}.

The enforcer and curators can---to some extent---be malicious, meaning they can deviate arbitrarily from the protocol.  Short of a way to prevent curators from adding non-harmful content, however, it may be more accurate to model them as \emph{covert} adversaries, meaning adversaries that can deviate from the protocol but ``do not wish to be `caught' doing so''~\cite{aumann07tcc}.  This captures the imperative for robust transparency and auditing of curated content to ensure blocklists do not become a tool for surveillance or censorship.  We also focus only on preventing malicious enforcers from selectively enforcing on specific content for specific clients, as it is always possible for them to deny service to a client by ignoring their requests or not distributing an updated software binary (although this latter risk can be mitigated through the use of \emph{binary transparency}~\cite{CCS:FDPFSS14,chainiac,contour,wired-gbt}).

Auditors also operate in an honest-but-curious fashion. By design, privileged auditors have access to the database of harmful objects, and unprivileged auditors have access to all public data and any reports that they receive from users (containing, e.g., their objects and their interactions with the enforcer).

\begin{table*}[]
    \centering
    \input{tables/threat_space}
    \caption{Design space for on-device protections including the threat information status of harmful content (\eg known vs. novel) and the intended detection outcome (\eg blocking vs. reporting). Our proposal targets the upper-left quadrant, narrowly focusing on \emph{preventing} the distribution of \emph{known} harmful content. For illustrative purposes, we list potential applications of each design quadrant.}
    \label{table:design_space}
\end{table*}

\subsection{Design principles}
We believe a core purpose of on-device blocklisting should be to \emph{prevent} the distribution of \emph{known} harmful content. This should ideally occur in a way that is robust, privacy-preserving, transparent, and auditable. Our emphasis here is crucial. Prevention---via blocking or warnings---is markedly different from automated reporting-based strategies (see Table~\ref{table:design_space}). Prevention is applicable to a broad set of harmful content---particularly scams, phishing, and malware---where automated reporting provides no immediate protection. Known harmful content is distinct from novel harmful content (see Table~\ref{table:design_space}). Detection of the latter often relies on the use of machine learning classifiers, near-duplicate matching, or manual review---all of which necessitate direct access to content or additional safeguards against false positives. We outline the set of design principles necessary to safeguard on-device blocklisting for all the parties in our threat model below.

\def\requestprivacy{Request privacy\xspace}
\paragraph{\requestprivacy} A query to determine whether an object is harmful should not leak any information about the content. While bounded content leakage may be appropriate for scenarios where clients have historically maintained less stringent security and privacy expectations, the same is not true in end-to-end encrypted contexts that have existing, strict privacy assumptions. Note that this is consistent with the definitions of E2EE presented in Section~\ref{sec:related}. Concretely, given the universe of objects \objset, for all $\objone, \objtwo \in \objset$ the request messages produced by a client running $\harmfuldetect$ on $\objone$ and the request messages produced by a client running $\harmfuldetect$ on $\objtwo$ are computationally indistinguishable.

Violations of this principle include k-anonymity, differential privacy, and side-channel leakage that would otherwise allow an adversary to distinguish requests on $\objone$ and $\objtwo$.

\def\verdictprivacy{Verdict privacy\xspace}
\paragraph{\verdictprivacy} The enforcer and curators should not learn the result of $\harmfuleval$. This is critical for applications where possession of erroneously detected harmful content may have legal ramifications, or where there is external pressure---from governments or platforms---to turn $\harmfuleval$ into a tool for surveillance (\eg learning about a user's activities). Users may choose to share the outcome with auditors, but this is optional and thus does not violate this principle. Further, users should be able to share the outcome anonymously without linking this result back to their identity.

\def\blocklistprivacy{Blocklist privacy\xspace}
\paragraph{\blocklistprivacy} The user and client should not learn anything from response messages $\resp$ obtained in the course of running $\harmfuldetect$ on $\objectshort$ except what is revealed by $\harmfuleval(\objectshort, \resp)$.  In other words, response messages should reveal information about only the queried object and not any other objects on the blocklist.

This is necessary to protect the secrecy of other potentially sensitive blocklist elements. Sensitivity may arise due to embarrassment (\eg a prominent, compromised domain involved in phishing campaigns); concerns around evasion; or the possibility of blocklist elements revealing personally identifying information.

\def\resistfalsepositives{Resilient to false positives\xspace}
\paragraph{\resistfalsepositives} Any verdict check needs to be resilient to false positives; i.e. it should be very unlikely for a client to query on an object $\objectshort$ and receive response messages $\resp$ such that $\harmfuleval(\objectshort, \resp) = 1$ but $\objectshort$ is not on the blocklist. This includes random false positives as well as adversarially generated objects that induce false positives. This principle shapes the underlying technologies that are viable for blocklist membership checks (\eg Bloom or Cuckoo filters; perceptual hashes; \etc). 

\def\prooforigin{Proof of origin\xspace}
\paragraph{\prooforigin} Clients should obtain evidence that irrefutably identifies all curators involved in arriving at a blocklist decision. Failure to produce sufficient proof should result in $\harmfuleval$ failing. Users should be able to convincingly report this evidence (and potentially any associated object) to any third party (including auditors), in order to hold enforcers and curators accountable for erroneous blocklist entries. However, given the principle around preserving blocklist privacy for non-queried objects, any reporting of an entry’s origin may implicitly reveal that the client possessed the related object.

\def\proofconsistency{Proof of consistency\xspace}
\paragraph{\proofconsistency} Blocklist entries and enforcement should be globally consistent, in order to avoid the risk of selective enforcement of blocklist entries across geographic regions or on a per-device basis.  The contents of the blocklist may change over time, however, so we consider consistency only within some time-bounded \emph{interval} rather than over all time. Concretely, for two clients $\clientone$ and $\clienttwo$ interacting with the enforcer within the same time interval \entitystyle{T} on the same object $\objectshort$ and receiving respective response messages $\resp_1$ and $\resp_2$:

\[
\harmfuleval(\objectshort, \resp_1) = 
\harmfuleval(\objectshort, \resp_2).
\]

{\noindent}This, along with other protections, help minimize the risk of misuse or censorship---discussed in Section~\ref{sec:practical}---as it ensures that if any enforcer or curator is compelled to expand the scope of blocklist entries, this will be globally identifiable.

\def\proofaudit{Proof of auditability\xspace}
\paragraph{\proofaudit} Clients should obtain evidence that irrefutably identifies the entire set of blocklist entries in enforcement at a given time \entitystyle{T}, such that a privileged auditor can reproduce the set of blocklist entries in order to ensure the correctness of verdicts that a client received during that period.

\def\realtime{Near real-time\xspace}
\def\resistdos{Resistance to denial of service\xspace}

\begin{table*}[t]
\centering
\footnotesize
\input{tables/comparison_framework}
\caption{Comparison of existing on-device blocklist proposals according to our design principles. We use a full circle \fullcircle to indicate a principle is fully realized, a half circle \halfcircle for when bounded leakage of privacy occurs, and a cross \emptycircle for when a principle is not satisfied.}\label{table:comparison}
\end{table*}

\paragraph{\realtime} In order for blocklisting to be effective for environments such as web browsers, storage, comments, or messaging applications, clients must obtain verdicts in near real-time. Runtime overhead may arise due to computationally expensive cryptographic operations that are performed by the client or enforcer, or network latency. This principle excludes any pre-work conducted by a client or enforcer prior to initiating $\harmfuldetect$.

\paragraph{\resistdos}
Across $\harmfuldetect$ interactions for an object set $\objset$, the amortized computational costs necessary for a client to produce request messages (including spoofed requests), denoted $t_c(\cdot)$, should be proportional to the computational costs of the enforcer, denoted $t_e(\cdot)$.

\begin{equation*}
    \sum_{\objectshort\in\objset} t_c(\objectshort) \propto
    \sum_{\objectshort\in\objset} t_e(\objectshort)
\end{equation*}

{\noindent}This amortized cost excludes any pre-work performed by the enforcer.

\subsection{Comparison of existing proposals}
\label{sec:comparison}
We evaluate to what degree existing on-device blocklist strategies satisfy our design principles in Table~\ref{table:comparison}. Our comparison excludes proposals that do not depend on  blocklists---such as user-initiated reporting~\cite{liu2021fighting} or meta-data based detection strategies~\cite{pereira2020metadata}. As we show in Section~\ref{sec:design} and~\ref{sec:stateless-design}, our proposals, if implemented properly, would satisfy all of our design principles.

\paragraph{Transparency \& Auditing} As shown in Table~\ref{table:comparison}, no existing proposal considers \emph{\prooforigin}, \emph{\proofconsistency}, and \emph{\proofaudit} in their design. Each proposal assumes enforcers engage in an honest-but-curious manner. This fails to satisfy our threat model, where we assume that enforcers and curators may surreptitiously add invalid blocklist elements if there is no built-in transparency or auditing to detect anomalies.

\paragraph{Robustness} Most proposals are not robust to random or maliciously-crafted collisions and thus fail to satisfy \emph{\resistfalsepositives}. These proposals rely on detection via perceptual hashes. Significant prior research has demonstrated these algorithms are vulnerable to adversarial examples, as the algorithms were not designed with such threats in mind~\cite{hao2021s, struppek2021learning, jain2021adversarial, wang2021prototype, dolhansky2020adversarial, bai2020targeted}. Protocols that do achieve this principle include Kogan \etal~\cite{kogan2021private}---designed for phishing URLs---and Thomas \etal~\cite{thomas2019protecting} and Pal \etal~\cite{li2019protocols}---designed for passwords---which all use cryptographic hashes. Other hypothetical examples that would violate this principle include cryptographic hashes that are susceptible to attacker-generated collisions.

\paragraph{Privacy-preserving} Proposals diverge on whether the client or enforcer learns the final verdict, with the latter failing to satisfy \emph{\verdictprivacy}. Not all proposals achieve \emph{\requestprivacy}, often due to tradeoffs where the client leaks some partial information about content (or associated metadata) in order to reduce the blocklist search space and achieve near real-time performance. These trade-offs are important explorations and accompanied by robust discussions on limitations and risks; we merely note in our comparison that if adopted, they would negatively impact end-to-end encryption assumptions around content privacy. Two proposals rely on two non-colluding enforcers; if this assumption were invalidated (\eg a government compelled both enforcers to share data), it would no longer satisfy \emph{\requestprivacy}.

\paragraph{Performance} Exact hash matching proposals are able to satisfy \emph{\realtime} (for sufficiently small blocklists in contexts where homomorphic encryption is used), while near-duplicate matching cannot achieve this principle without partially leaking information about the object to reduce overhead. Proposals that rely on homomorphic encryption incur non-trivial runtime overhead on enforcers, and thus present challenges to \emph{\resistdos}.

%% file: tables/threat_space.tex
\newcolumntype{P}[1]{>{\centering\arraybackslash}p{#1}}
\newcolumntype{M}[1]{>{\centering\arraybackslash}m{#1}}

\begin{tabular}{p{1.3cm}|p{1.3cm}|p{6.8cm}|p{6.8cm}}
\toprule
  \multicolumn{1}{M{1.3cm}|}{\bf  Content type} &
  \multicolumn{1}{M{1.3cm}|}{\bf  Technology space} &
  \multicolumn{1}{c|}{\bf   Automated blocking purpose} &
  \multicolumn{1}{c}{\bf  Automated reporting purpose} \\ 
\midrule
Known harmful content
    & Hash-based
    & \begin{itemize}[leftmargin=*,topsep=0pt]
    \item Prevent visits to scam, phishing, and malware URLs.
    \item Prevent redistribution of child sexual abuse media.
    \item Prevent information cascades of violent extremism, misinformation, and disinformation.
    \end{itemize} 
    & \begin{itemize}[leftmargin=*,topsep=0pt]
    \item Disable accounts distributing scams, phishing, and malware.
    \item Disable accounts distributing or engaging with child sexual abuse media, and report to law enforcement.
    \item Disable accounts distributing or engaging with violent extremist content, and report to law enforcement.
    \item Disable accounts distributing misinformation or disinformation.
    \end{itemize} 
    \\
  \midrule  
Novel harmful content
    & Machine learning-based
    & \begin{itemize}[leftmargin=*,topsep=0pt]
    \item Prevent rapidly evolving scam, phishing, and malware campaigns.
    \item Prevent grooming and production of novel child sexual abuse media.
    \item Prevent rapidly evolving information cascades related to violent extremism, misinformation, and disinformation.
    \end{itemize}
    & \begin{itemize}[leftmargin=*,topsep=0pt]
    \item Disable accounts distributing scams, phishing, and malware and improve detection signals.
    \item Enable law enforcement to identify and recover at-risk children.
    \item Disable accounts distributing misinformation and disinformation and improve detection signals.
    \item Enable law enforcement to investigate novel violent extremist threats.
    \end{itemize}
    \\
    \bottomrule
\end{tabular}

%% file: tables/comparison_framework.tex
\begin{tabular}{>{\raggedright}p{2cm}|>{\raggedright}p{8cm}|cccc|c|ccc|cc|}
\multicolumn{1}{c}{Proposal} & Description & 
\rotatebox{90}{Privacy holds w/ corrupt enforcer} &
\rotatebox{90}{\requestprivacy} &
\rotatebox{90}{\verdictprivacy} &
\rotatebox{90}{\blocklistprivacy} &
\rotatebox{90}{\resistfalsepositives} &
\rotatebox{90}{\prooforigin} &
\rotatebox{90}{\proofconsistency} &
\rotatebox{90}{\proofaudit} &
\rotatebox{90}{\realtime} &
\rotatebox{90}{\resistdos} \\
\toprule
Apple CSAM protocol~\cite{appleneuralhash} & Exact perceptual hash matching with PSM. Threshold secret sharing reveals harmful content and verdict to an enforcer. &
    \fullcircle & % threat model
    \emptycircle & % content privacy
    \emptycircle & % verdict privacy
    \fullcircle & % blocklist privacy
    \emptycircle & % resist FPs
    \emptycircle & % proof of origin
    \emptycircle & % proof of consistency
    \emptycircle & % proof of audit
    \fullcircle & % real-time
    \fullcircle \\ % resistdos
\midrule
Hua \etal~\cite{hua2021increasing} & Near-duplicate perceptual hash matching with PSM and bounded content leakage. &
    \fullcircle & % threat model
    \halfcircle & % content privacy
    \fullcircle & % verdict privacy
    \fullcircle& % blocklist privacy
    \emptycircle& % resist FPs
    \emptycircle& % proof of origin
    \emptycircle& % proof of consistency
    \emptycircle& % proof of audit
    \fullcircle& % real-time
    \fullcircle\\ % resistdos
\midrule
Kogan \etal~\cite{kogan2021private} & Cryptographic hash matching with PSM using two non-colluding enforcers. Two-phase lookup leaks information on blocklist verdict. &
    \emptycircle & % threat model
    \halfcircle& % content privacy
    \halfcircle& % verdict privacy
    \fullcircle& % blocklist privacy
    \fullcircle& % resist FPs
    \emptycircle& % proof of origin
    \emptycircle& % proof of consistency
    \emptycircle& % proof of audit
    \fullcircle& % real-time
    \fullcircle \\ % resistdos
\midrule
Kulshrestha \etal~\cite{kulshrestha2021identifying} & Exact, or near-duplicate, perceptual hash matching using FHE. Protocol supports client or enforcer learning verdict, or degree of a match. We assume client-only verdicts. & 
    \fullcircle & % threat model
    \fullcircle& % content privacy
    \fullcircle& % verdict privacy
    \fullcircle& % blocklist privacy
    \emptycircle& % resist FPs
    \emptycircle& % proof of origin
    \emptycircle& % proof of consistency
    \emptycircle& % proof of audit
    \emptycircle & % real-time
    \emptycircle \\ % resistdos
\midrule
Li \etal~\cite{li2019protocols} & Cryptographic hash matching with PSM and partial leakage of metadata about the object (\eg username) &
    \fullcircle & % threat model
    \halfcircle & % content privacy
    \fullcircle & % verdict privacy
    \fullcircle & % blocklist privacy
    \fullcircle & % resist FPs
    \emptycircle & % proof of origin
    \emptycircle & % proof of consistency
    \emptycircle & % proof of audit
    \fullcircle & % real-time
    \fullcircle \\ % resistdos
\midrule
Riazi \etal~\cite{riazi2016sub} & Near-duplicate perceptual hash matching with two non-colluding servers. Relies on bounded leakage of content. &
    \emptycircle & % threat model
    \halfcircle & % content privacy
    \fullcircle & % verdict privacy
    \fullcircle & % blocklist privacy
    \emptycircle & % resist FPs
    \emptycircle & % proof of origin
    \emptycircle & % proof of consistency
    \emptycircle & % proof of audit
    \fullcircle & % real-time
    \fullcircle \\ % resistdos
\midrule
Singh \etal~\cite{singh2019robust} & Perceptual hash generated remotely by an enforcer using FHE. Results in enforcer obtaining a plaintext hash of object, along with verdict. Scheme designed for hash algorithm privacy, not client privacy. &
    \fullcircle & % threat model
    \emptycircle & % content privacy
    \emptycircle & % verdict privacy
    \fullcircle & % blocklist privacy
    \emptycircle & % resist FPs
    \emptycircle & % proof of origin
    \emptycircle & % proof of consistency
    \emptycircle & % proof of audit
    \emptycircle & % real-time
    \emptycircle \\ % resistdos
\midrule
Thomas \etal~\cite{thomas2019protecting} & Cryptographic hash matching with PSM and partial leakage of metadata about the object. &
    \fullcircle & % threat model
    \halfcircle & % content privacy
    \fullcircle & % verdict privacy
    \fullcircle & % blocklist privacy
    \fullcircle & % resist FPs
    \emptycircle & % proof of origin
    \emptycircle & % proof of consistency
    \emptycircle & % proof of audit
    \fullcircle & % real-time
    \fullcircle \\ % resistdos
\bottomrule
\end{tabular}

%% file: 04a-design.tex
\section{Time-Efficient On-Device Blocklisting}\label{sec:design}
We present a time-efficient, on-device blocklisting protocol that satisfies the design principles discussed in Section~\ref{sec:framework}. Rather than invent new cryptographic primitives, we intentionally use proven, off-the-shelf technologies to facilitate near-term deployability. We begin with a high-level overview of the protocol and the notation we use for cryptographic and other operations. We then detail each phase of our protocol and how they satisfy our principles.

\subsection{Overview}
Similar to many existing proposals, in order to ensure \emph{\requestprivacy} and \emph{\blocklistprivacy}, we opt for a simple private set intersection (PSI) protocol that enables a client to query an encrypted database of elements controlled by an enforcer without revealing the object queried, or the elements of the database~\cite{huberman1999enhancing}. Our protocol necessitates this lookup to occur for every object the client encounters. As previously noted, any use of k-anonymity (such as a hash prefix) would leak either partial information about an object, while any use of a pre-filtering step (such as a Bloom filter or Cuckoo filter) would create a side-channel where follow-on requests---or the lack thereof---would reveal the subset of objects the client possessed. Avoiding such leakage entails that we encrypt and distribute the entire blocklist to clients (\eg a hash table, or trie)---a limitation we discuss later. Current implementations of PSI are relatively efficient for both clients and servers~\cite{pinkas2014faster}, satisfying our \emph{\realtime} and \emph{\resistdos} principles. PSI also makes no assumptions about non-collusion (compared to multi-party computation solutions), satisfying our threat model that privacy hold for dishonest enforcers. Additionally, as PSI is adaptable in terms of which party learns the intersection, we opt for only the client learning the result, resulting in \emph{\verdictprivacy}.

Our main departure from existing protocols is to add additional transparency and auditability into blocklist verdicts. Here, we rely on cryptographic signatures attached to every blocklist element to uniquely and irrefutably identify the curator claiming the object is harmful, satisfying \emph{\prooforigin}. In the absence of a signature---or a sufficient number of signatures---a client will ignore any blocklist verdict. Additionally, the entire blocklist is captured in an append-only Merkle tree that a client checks upon initialization, thus satisfying \emph{\proofconsistency}. Separately, privileged auditors can re-create the entire blocklist database distributed to clients, satisfying \emph{\proofaudit}.

Finally, we rely on SHA-256 cryptographic hashes as a way to uniquely represent objects in a space-efficient manner for conducting PSI comparisons. Cryptographic hashes are by definition \emph{\resistfalsepositives} due to their collision resistance. We leave a deeper discussion of adherence to our design principles to the Appendix.

\subsection{Preliminaries}

\paragraph{Notation} We use $\secp\in\NN$ to denote the security parameter and $\usecp$ to denote its unary representation. Algorithms are randomized unless explicitly noted otherwise.  We use
$y \gets A(x; r)$ to denote running algorithm $A$ on inputs $x$ and randomness $r$ and
assigning its output to $y$. We use $y\randpick A(x)$ to denote $y\gets A(x; r)$ for uniformly random $r$.

\paragraph{Signing}  We define a signature scheme $\sig$ according to the following three algorithms:
(1) the key generation algorithm $(\pk, \sk) \randpick \keygen(\usecp)$ takes in the security parameter and outputs a keypair consisting of a public and private key; 
(2) the signing algorithm $\sigma \randpick \sign(\sk, M)$ takes in the private key and a message and outputs a signature; and
(3) the (deterministic) verification algorithm $0/1 \gets \verify(\pk, M, \sigma)$ takes in the public key, the message, and the signature and outputs 1 if it believes the signature was produced by the party corresponding to that public key and 0 otherwise.  Such a signature scheme needs to be correct and to satisfy the standard notion of unforgeability (i.e. EUF-CMA security).

\paragraph{Encryption} We define a symmetric encryption scheme $\se$ according to the following three algorithms:
(1) the key generation algorithm $\sk \randpick \keygen(\usecp)$ takes in the security parameter and outputs a secret key; 
(2) the encryption algorithm $C \randpick\enc(\sk, M)$ takes in the secret key and a message and outputs a ciphertext; and 
(3) the (deterministic) decryption algorithm $M \gets\dec(\sk, C)$ takes in the private key and a ciphertext and outputs the underlying message.

\paragraph{Hashing} We define two hash functions: $H$ that maps to the message space of the signature and encryption schemes and $H_\key$ that maps to the space of possible encryption keys (more precisely, we could view $H_\key$ as the HKDF key derivation function). These hash functions are agreed upon by all parties and used consistently throughout our protocol.

\paragraph{Merkle trees}  Finally, for a Merkle tree $T$ we consider the following four deterministic algorithms:
(1) $\pi \gets \proveincl(T, \entry)$ proves the inclusion of entry in the tree;
(2) $0/1 \gets \verincl(h_{\mathsf{root}, T}, \entry, \pi)$ verifies the inclusion proof for entry with respect to the root hash of the tree (which acts as a commitment to its contents); 
(3) $\pi\gets \provecons(T, L_\mathsf{old}, L_{\mathsf{new}})$ proves that the tree of size $L_\mathsf{new}$ has been obtained only by adding entries to the tree of size $L_\mathsf{old}$; and
(4) $0/1\gets \vercons(h_{\mathsf{old}}, h_{\mathsf{new}}, \pi)$ verifies the consistency of the new and old tree roots.

\subsection{Protocol}
Following the conventions of Kamara et al.~\cite{kamara_outside_2022}, we separate our protocol into distinct phases: (1) definition, in which the curator and enforcer agree on the set of harmful content and take the necessary steps to place them under enforcement; (2) detection (corresponding to $\harmfuldetect$), in which the client and enforcer interact to identify objects that may be harmful; (3) evaluation (corresponding to $\harmfuleval$), in which the client confirms that the object is agreed to be harmful; (4) enforcement, in which the client takes action to block the harmful content; and (5) appeal, in which a user can dispute the accuracy of harmful verdicts.  We add two new phases: (6) audit, in which external, privileged auditors perform checks to ensure that the curator and enforcer are not enforcing on objects outside of their stated scope, and (7) update to support the addition or revocation of blocklist elements.

\paragraph{Definition} The definition of harmful content is made independently by a set of curators, uniquely identified by $j$, who we assume are in possession of some signing key pair $(\pk_j, \sk_j)$ and a set of harmful objects \objset.  Each curator maintains an append-only database with a monotonically increasing identifier that uniquely identifies a harmful object and its associated hash. In particular, for a harmful object $\obj_i$ a curator does the following:

\begin{itemize}[noitemsep]
\item Assigns it the identifier $\idx_i$.
\item Computes the object's hash $h_i \gets H(\obj_i)$.
\item Computes a signature as $\sigma_{j,i} \randpick \sign(\sk_j, h_i)$.
\end{itemize}

Each curator $j$ makes the set $\{\idx_i, h_{i}, \sigma_{j,i}\}_{i\in[N]}$  available to enforcers, and makes its public key $\pk_j$ available publicly.  Notably, it shares the raw content only in response to requests made by any privileged auditors. Enforcers, clients, users, and unprivileged auditors do not need access to the raw content of objects in the database.

 We also assume that each enforcer is in possession of some signing keypair, denoted $(\pk_E, \sk_E)$ and a blinding value $B$ that is randomly selected from $\FF_p^*$ for some prime $p$.  As an enforcer receives different sets of harmful objects from different curators, it can make its own decisions about which objects to put under enforcement; e.g., it might enforce only on objects that a sufficient number of curators agree are harmful.  According to this policy, the enforcer can select a subset of $M$ hashes that it opts to enforce on.  It then does the following for each element $i \in [M]$:

\begin{itemize}[noitemsep]
\item Computes a blinded object hash as $C_i \gets H(h_{i}^B\| \textrm{``id''})$.
\item Computes a key hash $h_{k,i}\gets H_\key(h_i^B \| \textrm{``key''})$.
\item For each curator $j$, computes an encrypted curator signature as $C_{\sigma,j,i} \randpick \enc(h_{k,i}, \sigma_{j,i})$.
\end{itemize}

It then distributes these to clients as a key-value store $C_i \mapsto \{C_{\sigma,j,i}\}_j$ and makes its public key $\pk_E$ available publicly.  As discussed previously, the signatures support the \emph{\prooforigin} property, and the blinding and encryption of both the object hash and its signature support \emph{\blocklistprivacy}.  Indeed, if the enforcer provided clients with unencrypted signatures, these might reveal information about $h_i$ or even allow the client to learn it entirely.

In order to ensure consistency, the enforcer should also provide each client with proof that it is seeing the same hashes under enforcement as all other clients at the same point in time.  To do this, the enforcer commits to the tuples $\{C_i, \{C_{\sigma,j,i}\}_j\}_{i\in[M]}$  via a hash $h_\mathsf{DB}$. It then appends this hash to a Merkle tree $T$ and provides the client with a \emph{checkpoint} $\chkpt = (h, L, t, \sigma)$ consisting of the root hash $h$ of $T$, its size $L$, a timestamp $t$, and a signature $\sigma\randpick \sign(\sk_E, (h, L, t))$, along with an inclusion proof $\pi\gets\proveincl(T, h_\mathsf{DB})$ proving that $h_\mathsf{DB}$ is the rightmost entry in the Merkle tree represented by the checkpoint.  To guarantee to the client that it is seeing the same view of this Merkle tree as other clients (i.e., is not subject to a \emph{split-view} attack), the enforcer can \emph{gossip} the checkpoint in some way; e.g., it can post the checkpoint to a public bulletin board and have the client retrieve it from there, or can have an external \emph{witness} or set of witnesses attest to the consistency of this checkpoint with all previous ones~\cite{syta2016keeping}.  Beyond this commitment to the contents of the dataset, the enforcer could also go one step further by including a commitment to the software binary in $h_\mathsf{DB}$, and thus achieve a degree of binary transparency (if an alternative method were not already in place).

Upon receiving the database values, the checkpoint $\chkpt = (h, L, t, \sigma)$, and the inclusion proof $\pi$, a client first checks that $\verify(\pk_E, (h, L, t), \sigma) = 1$.  If this check does not pass then it rejects the rest of the data from the enforcer.  If the checkpoint contains additional signatures from witnesses, then the client checks these signatures as well.  If all signature checks pass then the client forms the database and performs the same hashing as the enforcer to reconstruct the hash $h_\mathsf{DB}$.  It then checks that $\verincl(h, h_\mathsf{DB}, \pi) = 1$, and again rejects the data from the enforcer if this check does not pass.

In terms of storage cost, a client needs to permanently retain the database $C_i \mapsto \{C_{\sigma,j,i}\}$ throughout its operation, while $\chkpt$ and $\pi$ need not persist after verification of the database. Downloading the database from the enforcer is a one-time initialization cost, outside of periodic updates to accommodate new elements or revocation (discussed later). As such, there are no real-time constraints for the initialization process.

\paragraph{Detection}  In order for a client to detect if a given piece of content, denoted $\obj$, is under enforcement, it needs to determine if its local database contains the value $C = H(H(\obj)^B \| \textrm{``id''})$.  The client does not know $B$, however, so cannot compute this directly, and it cannot ask the enforcer to compute $C$ directly as this would involve revealing $H(\obj)$.  We thus have the client and enforcer engage in the Diffie-Hellman private set intersection (PSI) protocol, which proceeds as follows in a group of some prime order $p$: first, the client computes $A \randpick \FF_p^*$ and sends $\req \gets H(\obj)^A$ to the enforcer.  The enforcer then sends $\resp \gets \req^B$ back to the client, who computes $C' \gets \resp^{1/A}$.  This value is such that $C' = ((H(\obj)^A)^B)^{1/A} = H(\obj)^B$, and the client can thus compute $C = H(C'\| \textrm{``id''})$ and check if it is in the database as needed.

To prevent clients from quickly iterating through all possible objects, and thus learning the contents of the blocklist, it is important for the enforcer to \emph{rate limit} $\harmfuldetect$ interactions, for example by using the identity of the client or by requiring requests to be accompanied by a proof of computation.  We view the exact rate limiting mechanism as context-dependent and thus out of the scope of this paper.

\paragraph{Evaluation} If the value $C$ is not in the database, then the client concludes it is not harmful content and processes $\obj$ as normal.  If $C$ is present, before any enforcement action, the client verifies the signatures associated with $C$.  To do this, the client computes $K \gets H_\key(C'\| \textrm{``key''})$ and retrieves the encrypted values $\{C_{\sigma,j}\}_j$ from the database.
It then computes $\sigma_j \gets \dec(K, C_{\sigma,j})$ and runs $\verify(\pk_j, H(\obj), \sigma_j)$ for all $j$. If any of these verification checks fail, the client should treat the content as benign. The number of signatures that a client needs to satisfy this verification process is context-dependent, linearly increasing the storage cost. We discuss potential implementation optimizations, such as fixed-size multi-signatures, later in Section~\ref{sec:practical}.

\paragraph{Enforcement} If all signatures are valid, the client can now be sure the content is under enforcement, as agreed by the curators and the enforcer (demonstrated implicitly through the object's Merkle inclusion proof).  At this point it blocks the content or otherwise warns the user. The final enforcement mechanism is context-dependent and outside the scope of this paper, but automated dissemination of the verdict or content would violate \emph{\verdictprivacy} or \emph{\requestprivacy} respectively.

\paragraph{Appeal} In order for users to dispute enforcement decisions, they can share $\obj$ and the curator signatures $\{C_{\sigma,j}\}$ publicly or via an unprivileged auditor who conceals the user's identity (\eg a journalist or court). Here, the signatures provide irrefutable proof that these curators are actively blocklisting the object, while the raw $\obj$ is necessary for any manual or public review on whether or not the object is indeed harmful. Similarly, if a signature $C_{\sigma,j}$ is invalid but still included in the client's database, the user can alert the enforcer or curator. However, doing so implicitly reveals that the user had access to $H(\obj)$ in order to obtain $H(\obj)^B$ and decrypt the signature. These privacy concerns can be avoided through privileged auditing (discussed shortly), which avoids the necessity of clients sharing their objects, hashes, or signatures. Alternatively, users can rely on an anonymization service like Tor.

\paragraph{Audit} There are several aspects of the protocol described above for which an auditor might want to ensure that the enforcer and curators are not misbehaving. First, an unprivileged auditor might want to ensure that all clients see the same objects under enforcement.  To do this, an auditor can periodically obtain all checkpoints $(\chkpt_1, \ldots, \chkpt_\ell)$ of the Merkle tree maintained by the enforcer, according to the gossip protocol (e.g., they can retrieve the checkpoints stored with a set of witnesses or on a public blockchain).  It can then check the consistency of these checkpoints by, for example, asking the enforcer for a consistency proof $\pi$ between the tree at size $L_i$ and the tree at size $L_\ell$ for all $i$ and then checking that $\vercons(\chkpt_i, \chkpt_\ell, \pi) = 1$.  Likewise, the auditor can check that the timestamps contained in the checkpoints follow some acceptable policy concerning how often the enforcer can update its dataset.  As one consequence, this allows the auditor to detect \emph{oscillation attacks}~\cite{meiklejohn2020gossip} in which the enforcer might try to rapidly switch between different objects under enforcement for different types of clients.  Obtaining consistency proofs provides information only about the internal and leaf hashes in the tree, which means that an enforcer should reasonably be able to make access to the Merkle tree public and thus make this type of auditing possible even for an unprivileged auditor.

A privileged auditor with access to the underlying objects could go one step further by checking the actual contents of the database, both in terms of making sure that the values it contains correspond with the values in the Merkle tree, and that these objects really do fall into the category of harmful content.  The latter type of check is out of the scope of this paper, but for the former type recall that each leaf hash in the Merkle tree represents a set of values $\{C_i, \{C_{\sigma, j, i}\}_j\}_{i\in[M]}$.  The auditor can request each of these sets of values and check that they do hash to the appropriate leaf hash.  It can then request the set of hashes underlying the $C_i$ values, along with the set of objects underlying those hashes.  To ensure that these are the right objects for this set, the auditor can re-run the checks of the client; i.e. it forms the encryption key $K_i\gets H_\key(H(\obj_i)^B \| \textrm{``key''})$ for each object $\obj_i$, decrypts $\sigma_{j,i}\gets \dec(K_i, C_{\sigma, j, i})$ for all $j$, and checks that $\verify(\pk_j, H(\obj_i), \sigma_{j, i}) = 1$ for all $j$.

\paragraph{Update}
Due to the dynamic nature of blocklists---for example, due to the remediation of compromised websites involved in phishing, the discovery of new malware binaries, or successful appeals---our protocol needs to support the addition and revocation of blocklist elements. Our protocol supports updates via enforcers distributing partial (or complete) modifications to the key-value store $C_i \mapsto \{C_{\sigma,j,i}\}$, which also entails updating $\chkpt$ and providing clients with a new inclusion proof $\pi$. In order to prevent an enforcer from re-using revoked signatures from curators, rather than distributing a list of revoked signatures, there are two main options.  First, curators and enforcers could update their signing keys on a regular interval known to clients.  Clients would then perform signature verification using the keys for a given period, treating all signatures computed using keys in previous periods as stale and invalid.  This strategy mirrors that of short-lived certificates~\cite{topalovic2012towards}.  Second, curators could make public a timestamp $\entitystyle{T}_j$ such that signatures are valid for only some known validity period after this timestamp, and then produce signatures using $\sigma_{j,i}\randpick\sign(\sk_j, (\entitystyle{T}_j, h_i))$.  Again, clients would then check the freshness of the timestamp in addition to performing signature verification, and treat as invalid all signatures that use an outdated timestamp.

%% file: 04b-rlwe-design.tex
\section{Space-Efficient On-Device Blocklisting}\label{sec:stateless-design}
We present an alternative, space-efficient on-device blocklisting protocol for scenarios that do not need \emph{\realtime} and \emph{\resistdos}, but that otherwise satisfies all our other design principles.

\subsection{Overview}
In order to support scenarios where an entire blocklist cannot fit on-device, our space-efficient protocol relies on fully homomorphic encryption to privately obtain a partition of the blocklist at the time of $\harmfuldetect$, without revealing the partition queried, or learning about any other entries in the blocklist—as previously proposed by Kulshrestha and Mayer~\cite{kulshrestha2021identifying}. This satisfies \emph{\requestprivacy}, \emph{\verdictprivacy}, and \emph{\blocklistprivacy}. As with our time-efficient protocol, we ensure all entries present in the resulting partition (as well as all unobserved partitions) are cryptographically signed and present in a Merkle tree that can be reconstructed by an external auditor, thus satisfying \emph{\prooforigin}, \emph{\proofconsistency}, and \emph{\proofaudit}. We again rely on SHA-256 cryptographic hashes to ensure \emph{\resistfalsepositives}. The computational overhead of fully homomorphic encryption fails to satisfy \emph{\realtime} and \emph{\resistdos}. We discuss scenarios where these principles may not be needed later in Section~\ref{sec:practical}.

\subsection{Preliminaries}
\paragraph{FHE} In addition to the cryptographic primitives used in Section~\ref{sec:design}, our space-efficient protocol makes use of a symmetric fully homomorphic encryption (FHE) scheme.  We define this according to the following six algorithms:
First, the key generation algorithm $\sk\randpick\keygen(\usecp)$ takes in the security parameter and outputs a secret key.
Second, the encryption algorithm $C\randpick\enc(\sk, M)$ takes in the key and a message and outputs a ciphertext.
Third, the homomorphic \emph{addition} algorithm $C\randpick\add(\{C_i\})$ takes in encryptions of plaintexts $M_1, \ldots, M_n$ and outputs an encryption of $M_1 + \cdots + M_n$.
Fourth, the homomorphic \emph{multiplication} algorithm $C\randpick\mult(\{C_i\})$ takes in encryption of plaintexts $M_1, \ldots, M_n$ and outputs an encryption of $M_1\times \cdots\times M_n$.
Fifth, the \emph{absorb} algorithm $C'\randpick\absorb(C, M')$ takes in an encryption of some plaintext $M$ and a plaintext $M'$ and outputs an encryption of $M\oplus M'$ for some operation $\oplus$.
Sixth and finally, the decryption algorithm $M\gets\dec(\sk, C)$ takes in the secret key and a ciphertext and outputs the underlying message.

\subsection{Protocol}

\paragraph{Definition} The way curators define harmful content is the same as in the previous protocol.  This means a curator $j$ makes available to enforcers a set $\{\idx_i, h_{i}, \sigma_{j,i}\}_{i\in[N]}$, where $h_i\gets H(\obj_i)$ is the hash of the harmful object $\obj_i$ and $\sigma_{j,i}$ is a signature on this hash under the curator's key.

Also following the previous protocol, for every element $i$ enforcers compute a blinded object hash $C_i\gets H(h_i^B \| \textrm{``id''})$ (for some blinding value $B\randpick \FF_p^*$), a key hash $h_{k,i}\gets H_\key(h_i^B \| \textrm{``key''})$, and encrypted curator signatures $C_{\sigma,j,i}\randpick\enc(h_{k,i}, \sigma_{j,i})$.  Unlike in the previous protocol, however, the enforcer does not distribute these objects to the client.  Instead, it sorts each blinded object hash into a \emph{bucket} according to its $k$-bit hash prefix. It then identifies the maximum bucket size $S$, and pads each bucket with null hash values to ensure that they all have the same size $S$.  This provides the server with a collection of buckets $B_1, \ldots, B_{2^k}$, where each entry in a bucket is a pair consisting of a blinded object hash with the appropriate hash prefix and a set of encrypted curator signatures.

In order to ensure consistency, the enforcer should also provide each client with proof that it is seeing the same hashes under enforcement as all other clients at the same point in time.  To do this, the enforcer commits to the contents of each bucket $B_\ell$ by computing $h_\ell\gets H(C_1 \| C_{\sigma, 1, 1} \| \cdots \| C_{\sigma, n, 1} \| \cdots \| C_S \| C_{\sigma, 1, S} \| \cdots \| C_{\sigma, n, S})$ for all $(C_i, \{C_{\sigma,j,i}\}_j)$ in $B_\ell$.  It then commits to the collection of buckets, representing the entire database, by computing $h_{\mathsf{DB}}\gets H(h_1\| \cdots \| h_{2^k})$.  As in the previous protocol, it then appends this hash to a Merkle tree $T$, which results in it having a new checkpoint $\chkpt$.  It could also fold a commitment to the software binary into this hash if there were not already a method in place for achieving binary transparency.

\begin{table*}[t]
\footnotesize
\centering

\begin{tabularx}{0.9\textwidth}{X|ccc|lccc}
\toprule
                        &  \multicolumn{3}{c|}{\bf Time-efficient}                      & \multicolumn{4}{c}{\bf Space-efficient}                                                    \\
                        
% Header
    &
    \it Pixel 2XL &
    \it Pixel 6 Pro &
    \it Bandwidth &
    \it Parameter &
    \it Pixel 2XL &
    \it Pixel 6 Pro &
    \it Bandwidth \\
\midrule
% Row 1
    \multirow{2}{*}{\bf Query} &
    \multirow{2}{*}{425 $\mu$s} &
    \multirow{2}{*}{100 $\mu$s} &
    \multirow{2}{*}{32 bytes}     &
    $N=4096$ &
    7.99 ms   &
    2.99 ms     &
    92.9 KB   \\
% Row 2    
    &
    &
    &
    & 
    $N=8192$ &
    20.4 ms   &
    6.85 ms     &
    432.7 KB  \\
    \midrule
% Row 3
    \multirow{2}{*}{\bf Verification} &
    \multirow{2}{*}{257 $\mu$s} & 
    \multirow{2}{*}{69 $\mu$s}  & 
    \multirow{2}{*}{32 bytes}   &
        $N=4096$ &
    8.35 ms   &
    2.57 ms     & 
    10.2 KB   \\
% Row 4
    &                           
    &  
    &
    & 
    $N=8192$ & 
    36.7 ms   & 
    10.2 ms     & 
    20.5 KB   
    \\
\bottomrule
\end{tabularx}
\caption{Client-side computation and bandwidth costs to determine the blocklist status of a single object.}\label{table:bandwidth}
\end{table*}

\paragraph{Detection} In order for a client to detect if a given piece of content, denoted $\obj$, is under enforcement, it needs to determine if the enforcer's database contains the value $C = H(H(\obj)^B \| \textrm{``id''})$.  As in the previous protocol, we start by having the client and enforcer run the Diffie-Hellman private set intersection protocol in order for the client to learn this value without revealing $H(\obj)$ to the enforcer.

Next, the client and enforcer interact to determine if $C$ is in the enforcer's database.  This interaction largely follows the approach taken by Kulshrestha and Mayer~\cite{kulshrestha2021identifying}: first, the client computes the $k$-bit hash prefix of $C$ and converts this to an integer value $\alpha$.  It then forms a vector $v$ such that $v_j = 1$ if $j = \alpha$ and $v_j = 0$ otherwise, samples a secret key $\sk\randpick\fhe.\keygen(\usecp)$, and computes $C_j\gets\fhe.\enc(\sk, v_j)$ for all $j$.  After receiving this ciphertext vector, the enforcer computes $A_j\randpick\fhe.\absorb(C_j, B_j)$ for all $j$ and $A\randpick\fhe.\add(\{A_j\})$, sending $A$ back to the client.  By how $v$ is defined, this is an encryption of the bucket $B_\alpha$.  It also sends back all the bucket commitments $\coms\gets \{h_\ell\}_{\ell=1}^{2^k}$, an inclusion proof $\pi\gets\proveincl(T, h_{\mathsf{DB}})$, and an appropriately gossiped checkpoint $\chkpt$ containing the root of $T$ (alternatively the client can obtain this checkpoint from a separate source, as discussed for the previous protocol).  The client then computes $B_\alpha\gets\dec(\sk, A)$ and checks if $(C, \cdot)$ is in this list.

\paragraph{Evaluation} If the value $C$ is not in $B_\alpha$, then the client concludes it is not harmful content and processes $\obj$ as normal.  If $C$ is present, the client next verifies the proof of consistency provided by the enforcer and the signatures associated with $C$.  To check for consistency, the client first checks the signature(s) in $\chkpt$; if any of these checks fail then the client treats the content as benign.  It then recomputes the commitment to $B_\alpha$ by forming $h_\alpha\gets H(C_1 \| C_{\sigma, 1, 1} \| \cdots \| C_{\sigma, n, 1} \| \cdots \| C_S \| C_{\sigma, 1, S} \| \cdots \| C_{\sigma, n, S})$ for all $(C_i, \{C_{\sigma, j, i}\}_j)$ in $B_\alpha$.  It then checks that $h_\alpha\in\coms$ and that $\verincl(\chkpt, h_\mathsf{DB}, \pi) = 1$ for $h_\mathsf{DB}\gets H(\coms)$, again treating the content as benign if either of these checks fail.  To check the signatures, the client computes $K \gets H_\key(H(\obj)^B \| \textrm{``key''})$ and retrieves the encrypted signatures $\{C_{\sigma,j}\}$ from the bucket pair containing $C$.  It then computes $\sigma_j \gets \dec(K, C_{\sigma,j})$ and runs $\verify(\pk_j, H(\obj), \sigma_j)$ for all $j$.  Again, if any of these checks fail then the client should treat the content as benign.

\paragraph{Enforcement, Appeal, Audit}
These phases are all the same as in the previous time-efficient protocol.

\paragraph{Update}
Unlike in the time-efficient protocol, updating the blocklist does not involve communicating with the client, who stores no data.  Nevertheless, updating the blocklist still produces a new checkpoint $\chkpt$ and the enforcer must ensure that this gets gossiped before the new blocklist can be used.

%% file: 05a-evaluation.tex
\section{Evaluation}\label{sec:evaluation}

We provide micro-benchmarks on the storage costs, bandwidth, and processing costs of both our time-efficient and space-efficient protocol. Summaries of the client-side computation and bandwidth for both protocols are available in Table~\ref{table:bandwidth} and the server-side computation costs for our space-efficient protocol in Table~\ref{table:server_space_efficient}.

\subsection{Implementation}
We implemented our time-efficient protocol using Rust, with Rust's AES\_GCM library~\cite{aes_gcm_rust} for symmetric encryption; Dalek Cryptography’s 25519 curve~\cite{dalek_curve25519} and ed25519~\cite{dalek_ed25519} for signing, verification, and blinding; and SHA256 for hashing. For the PIR component of our space-efficient protocol, we used a modified version of SealPIR~\cite{SealPIR} written in C++. We note that SEAL uses the BFV cryptosystem, which can encode $\approx \lfloor(N * 20) / 8\rfloor$ bytes per index, where N is the security parameter. We test with $N=4096$ and $N=8192$.\footnote{Prior works use $N=2048$~\cite{SealPIR, mulpir, kulshrestha2021identifying}. SEAL has since been updated, requiring a minimum of $N=4096$ as a security parameter.} 

We benchmarked all client-side computation on two mobile devices with varying capabilities: a Pixel 6 Pro (released April 2022), and a Pixel 2XL (released October 2017). For server-side computation, we used a 96-core Intel Xeon 2.0GhZ server with 177G of RAM. We benchmarked Rust components with Criterion~\cite{criterion} and C++ components with Google's Benchmarking suite~\cite{google_benchmark}, with a minimum of 200 iterations. We omit network latency which will vary by provider, but report the bandwidth costs per request. For brevity, we omit benchmarks not on a critical path such as the client’s verification of inclusion proofs (performed on initialization) or signature verification (performed only in the event of a match).

\subsection{Time-efficient protocol performance}

\paragraph{Storage cost} For our time-efficient protocol, a single blocklist element consists of a blinded hash (256 bits); an ed25519 signature (512 bits) per curator; and, as AESGCM is length preserving, two bytes for a nonce (16 bits). As a result, each entry’s size is roughly $(256 +16 + (512*j))/8$ bytes, where $j$ is the number of curator signatures per entry. Assuming only one curator, a blocklist of 50K entries would need $\sim $4.9MB of storage. A blocklist of 1M entries would need $\sim$98MB.

\paragraph{Client computation \& bandwidth} We benchmarked the time necessary for our test device to compute a SHA256 hash, and then perform a scalar multiplication with a random value on the resulting hash.\footnote{We compute our hash over an entity of 32 bytes. This computation will scale linearly for larger objects such as images or videos, but may be amortized if  already performed as part of sending or receiving the object.} On a Pixel 6, this combined blinding operation took an average of 100$\mu$s and on a Pixel 2XL, 425$\mu$s.

We also measured the time it took a client, upon receiving a blinded hash from the server, to expand the point, invert their key, and perform a scalar multiplication. On a Pixel 6, this took an average of 69$\mu$s and on a Pixel 2XL, 257$\mu$s. In terms of bandwidth, a query needs just 32 bytes.

\paragraph{Server computation \& bandwidth} To respond to a query, the server must decompress the edwards point and perform a single scalar multiplication on the query. Our server took an average of 76$\mu$s per message. In terms of bandwidth, a response needs just 32 bytes. In practice, we expect that optimizations such as batch processing would provide significant speed improvements.

Taken as a whole, any delay to sending or receiving a message in our time-efficient protocol will be dominated by round-trip network latency. In practice, computation and bandwidth costs appear within the tolerance of many types of devices.

\subsection{Space-efficient protocol performance}

\paragraph{Storage cost of blocklist} Storage costs in our space-efficient protocol are transient (persisting only until after verification), and proportional to the size of each bucket. For our implementation, this results in a plaintext size of 10,240 bytes ($N=4096$), and 20,480 bytes ($N=8192$). 

\paragraph{Client computation \& bandwidth} Client computation and bandwidth in SealPIR are independent of the number of entries.\footnote{So long as a query can fit within a single ciphertext, translating to a blocklist of fewer than $2^{25}$ elements in our configuration.} On a Pixel 6, composing a query cost 2.99ms for a security parameter of $N=4096$. On a Pixel 2XL, this cost 7.99ms. This is an order of magnitude more than the blinding request, which our space-efficient protocol must also perform. In terms of bandwidth, our space-efficient protocol needs the client to send a 92.9KB message to the server.

Upon receiving a response, the Pixel 6 spent an additional 2.57ms on decryption, while the Pixel 2XL spent 8.35ms for $N=4096$. The combined time to generate a request, send a multi-KB query, and process a response is why we do not consider our space-efficient protocol to be \emph{\realtime}.

\paragraph{Server computation \& bandwidth} Server costs in SealPIR scale as a function of the number of blocklist entries. Assuming only one curator, a single blocklist element needs 98 bytes. This means that a single index can store $10,240/98 = 104$ (for $N=4096$) entries, and $208$ entries for $N=8192$. As a result, for a given bit prefix S, the maximum number of entries we can store is $2^s*104$ (N=4096) or $2^s*208$ (N=8192).

The server’s computation can be parallelized to reduce response delay, but the server-side costs we report are non-negligible (see Table~\ref{table:server_space_efficient}).
For a blocklist of roughly 54K elements, a server must expend 183ms of computation per request. For a blocklist of roughly 870K elements, the server must expend 1.4s per request. For bandwidth, a response needs 10.2KB ($N=4096$), in addition to the 32 bytes needed for the blinding operation. 
This computation imbalance means our space-efficient protocol does not satisfy \emph{\resistdos}. As a whole, such costs may be out of reach in the near-term for both servers and clients.

\begin{table}[t]
\setlength{\tabcolsep}{10pt}
\footnotesize
\centering
\begin{tabularx}{\columnwidth}{Xll|ll}
\toprule
\multicolumn{1}{l|}{}    & \multicolumn{2}{c|}{$N=4096$}                               & \multicolumn{2}{c}{$N=8192$}                               \\
\multicolumn{1}{r|}{\bf S}  & \multicolumn{1}{c}{\bf Elements} & \multicolumn{1}{c|}{\bf Time (ms)} & \multicolumn{1}{c}{\bf Elements} & \multicolumn{1}{c}{\bf Time (ms)} \\ \midrule

\multicolumn{1}{r|}{6}  & 6,826                       & 53.0                           & 13,653                      & 262                           \\
\multicolumn{1}{r|}{7}  & 13,653                      & 76.3                           & 27,306                      & 368                           \\
\multicolumn{1}{r|}{8}  & 27,306                      & 118                            & 54,613                      & 567                           \\
\multicolumn{1}{r|}{9}  & 54,613                      & 183                            & 109,226                     & 853                           \\
\multicolumn{1}{r|}{10} & 109,226                     & 290                            & 218,453                     & 1,344                          \\
\multicolumn{1}{r|}{11} & 218,453                     & 482                            & 436,906                     & 2,121                          \\
\multicolumn{1}{r|}{12} & 436,906                     & 802                            & 873,813                     & 3,571                          \\
\multicolumn{1}{r|}{13} & 873,813                     & 1,400                           & 1,747,626                    & 6,099                          \\
\multicolumn{1}{r|}{14} & 1,747,626                    & 2,492                           & 3,495,253                    & 10,786                         \\
\multicolumn{1}{r|}{15} & 3,495,253                    & 4,599                           & 6,990,506                    & 19,604                         \\
\bottomrule
\end{tabularx}
\caption{Server-side computation cost for space-efficient protocol for a blocklist with a variable number of elements.}
\label{table:server_space_efficient}
\end{table}

%% file: 05b-deployment.tex
\section{Discussion and Practical Considerations}\label{sec:practical}
In light of the performance of our protocols, we discuss practical considerations---such as blocklist size, or where to enforce blocklisting---that will arise in real deployment scenarios.

\paragraph{Blocklist size} Our time-efficient protocol entails that clients store $C_i \mapsto \{C_{\sigma,j,i}\}$ locally on-device. As detailed in our evaluation in Section~\ref{sec:evaluation}, this space cost grows linearly with the number of blocklist elements and the number of curator signatures involved, and requires only 90MB for a million entries. 

We believe these storage requirements are practical. 
For context, at the time of writing, WhatsApp itself requires a minimum of 90MB of storage on-disk, not including user data or other resources. 
We also envision that storage would be amortized, with on-device blocklisting provided as a device-wide service, with multiple applications seeking protection from a common set of harmful content.

In addition, there are a number of ways one might work around storage constraints. For deployment scenarios with more than one curator, for example, the storage cost of signatures can be reduced by relying on fixed-size multi-signature schemes, like BLS~\cite{boneh2001short}. Finally, if harmful content's distribution frequency follows a Zipf-like distribution, it may be possible for enforcers to optimize blocklist sizes to include only the head of the distribution, at the cost of some false negatives in the tail, and measurement studies indicate that harmful content often have a very limited shelf-life~\cite{bursztein2019rethinking}.
This approach would apply equally well if harmful content could be ranked by its severity, with enforcement taking place on only the most severe instances.

\paragraph{Revocation \& signature validity window} As we discussed in Section~\ref{sec:design}, revocation relies on either a timestamp or a new key to prevent outdated blocklist elements from being enforced. One possibility is for all $j$ curators to independently manage the validity windows of their signatures $\{C_{\sigma,i}\}$, which may lead to inconsistent revocation timelines. Conversely, the enforcer might advertise a fixed time window for use in the construction of all signatures or determining when to rotate keys, but that necessitates a highly technical means of coordination across all $j$ curators---something that may be challenging in practice. This is true even more so if there are multiple, independent enforcers. Given the complexity of revocation in TLS environments~\cite{larisch2017crlite}, reducing the validity window of signatures is likely an area for future work.

\paragraph{On-receipt vs. on-send protections} Our protocols are agnostic to whether blocklisting is performed on-receipt or on-send (for storage, the equivalent of on-download or on-upload), and this decision is context dependent. Enforcing protections on-send, such as in a messaging or storage application, avoids the need to remove or otherwise silo harmful content on the recipient’s device or remote server, which is especially pertinent when possession of harmful content carries legal ramifications. It also makes it more clear how users could report content that they do not believe should be considered harmful.

Conversely, on-receipt protections are more suited to browsing or other delivery mechanisms with a potentially unknown origin of content, and might better match users’ mental models in more closely matching existing protections such as spam filters. While we explicitly consider misbehaving clients as out-of-scope for our threat model, in practice, malware and phishing campaign operators may send content via clients where blocklists are not enforced, thus requiring on-receipt detection even in messaging and storage contexts. Future work might address how content under transmission might self-prove it has undergone blocklist evaluation.

\paragraph{Blocking, warning, labeling, \& more}
Once a client encounters harmful content, in practice, it can employ a number of content moderation strategies ranging from outright blocking the content with no ability for users to override the decision; warning the user such that they can override the decision; or label the content with additional security information (\eg an untrusted source), while still displaying the content. Our intent in enumerating these options is to demonstrate that on-device blocklisting is merely a signal that can feed into more complex client enforcement logic.

\paragraph{Deniability \& forward secrecy of verdict results}
It may be that blocklisting verdicts need to be deleted immediately unless otherwise requested by the user. Having cryptographic proof that one was notified of flagged content could lead to discovery and punishment, especially under compelled decryption scenarios, which adds to the challenges of maintaining verdict secrecy and protecting against misuse for censorship or other reasons. 

\paragraph{Curator capabilities \& selection} 
We intentionally designed our protocol to separate the role of the enforcer---who handles all the technical infrastructure necessary for content serving, client updates, and responding to queries---from curators---who exclusively identify harmful content and provide a signature $C_{\sigma,j,i}$. This avoids the need for curators to support burdensome computational and systems costs, which could be substantial for larger-scale messaging systems, storage, or browser interactions. Additionally, our protocol needs no communication between individual curators, while also allowing enforcers to be flexible in which curators' threat intelligence they incorporate into blocklist verdicts. 

Finally, the protocol allows for clients to decide, based on the curator, if the verdict should be respected. This property is an important distinction from most blocklisting protocols---users often do not have uniform trust models, and may have varying trust in governments or private institutions to provide unbiased blocklists. 

%% file: 06-limitations.tex
\section{Limitations}\label{sec:limitations}

\paragraph{Dispute resolution} 
We make a number of assumptions about the powers of the enforcer, curators, auditors, and the general public to hold the system accountable and remediate failures. In particular, we assume that, if a non-harmful item is added to the database, a user can and will share that result with the enforcer or a third party (e.g, a reporter, a court), and that seeking redress will be effective in correcting the identified failure and altering the behavior of the enforcer and curators in the future. It is possible such assumptions do not hold, in which case the deployment of such a system will yield significant, albeit known, risks for censorship.  

\paragraph{Unintended expansion to other use cases} 
Some may take the adoption of anti-abuse tools in E2EE environments as proof that more complex analyses such as inter-user abuse and novel harmful content detection can be performed as well with similar levels of protection. We stress that any such expansions may violate the principles we outlined above (e.g., against false positives, or content and verdict privacy). For example, any system that automatically reports verdicts or content to outside parties would violate commonly held definitions of E2EE.

\paragraph{Censorship risks}
Our proposed protections may be deployed in an opt-in or opt-out manner, keeping in mind that the device should be viewed as an agent of the user. In an opt-in context, there is still the risk of a slippery slope in terms of unintended expansion (as discussed above) even for users who opt in. The protections may also be deployed to all devices, making it possible for a government or attacker to compel enforcement even for users who do not opt in. In an opt-out context, there is the risk to all users of having non-harmful content censored, with the only potential recourse being the ability to report.

\paragraph{Evasion \& modification} 
Malicious actors might leverage the system itself as an oracle to determine if their harmful content is blocked, and then change their behavior to evade the result. We note that this is a universal problem in most anti-abuse systems (e.g. an email spammer may send themselves a target message to determine if it will be blocked). We emphasize that this system---or any---cannot stop concerted attackers from sending harmful content to one another~\cite{horel_how_2018}, and that representatives from law enforcement have explicitly stated that the spread of illegal viral images, as well as the dissemination of known images to unsuspecting victims, are both of grave concern~\cite{robinsonlevy}.

\paragraph{Complexity \& bugs} 
Satisfying our design principles hinges on the correctness of any implementation. In practice, any additional code added to a system can increase risk of security vulnerabilities such as unintended code execution, side channels, or other vulnerabilities. 

%% file: 07conclusion.tex
\section{Conclusion}

In this paper, we presented principles necessary to ensure that on-device blocklisting is robust, privacy-preserving, transparent, and auditable and outlined two constructions---one that is time-efficient, and one that is space-efficient---that meet these principles. Our benchmarks show that our time-efficient protocol can operate on a variety of devices, while our space-efficient protocol remains a worthwhile direction as the underlying cryptographic primitives improve. In exploring this design space, we again emphasize that we take no position on whether or not the protocols discussed in this paper should be adopted in practice. We hope our work serves as a useful discussion point in the continued study of on-device abuse prevention---specifically for end-to-end encrypted environments---as it pertains to user expectations, improving cryptographic primitives, resistance to censorship and surveillance, and evolving privacy best practices. 

%% file: acks.tex
\section*{Acknowledgements}
We thank our anonymous reviewers, colleagues, and experts for their feedback and suggestions which helped to shape this work.

%% file: appendix.tex
\section*{Appendix}
\subsection*{Adherence to design principles}
\label{sec:rlwe-security}

Our time-efficient protocol satisfies all of our design principles, while our space-efficient protocol satisfies all properties other than the \emph{\realtime} and \emph{\resistdos} principles. We explain in detail why below.

\paragraph{\requestprivacy}  The pair of request messages for an object $\obj$ is of the form $(H(\obj)^r, \{C_j\})$ for some randomness $r$ and for $C_j\gets\fhe.\enc(\sk, v_j)$ for a bit $v_j\in\{0,1\}$.  For the former, there is always some randomness that is consistent with a request for another object; i.e. for $\obj$ there exists $\obj'$ such that $H(\obj)^r = H(\obj')^s$ for $s = r \cdot \mathsf{dlog}_{H(\obj')}(H(\obj))$, which is uniformly distributed.  These distributions are thus identical, so overall the request messages for different objects are computationally indistinguishable. For the latter, the IND-CPA security of the encryption scheme guarantees that encryptions of $0$s and $1$s are computationally indistinguishable.

\paragraph{\verdictprivacy} As the verdict is decided and acted upon locally by the client and the request messages do not reveal any information about the object (per \emph{\requestprivacy}), there is no way for the enforcer or curators to learn any information about the verdict.

\paragraph{\blocklistprivacy} In interacting with the enforcer, the client sees blinded object hashes and encrypted signatures.  By the security of the encryption scheme, these ciphertexts do not reveal any information about the signatures, and thus the underlying messages.  Likewise, by the security of the PSI protocol, the blinded object hashes do not reveal any information about the underlying object hashes.  The natural exception is that the client does learn information when it identifies a match in its database, but this necessitates it to know the underlying object. In practice, some form of rate limiting access to $\harmfuldetect$ interactions is necessary to prevent a client from enumerating all possible hashes---or a limited search for a set of objects---to determine their blocklist status.

\paragraph{\resistfalsepositives} A false positive means that a client found an object $\obj$ such that $C = H(H(\obj)^B \| \textrm{"id"})$ is in their database and $\verify(\pk_j, H(\obj), \dec(H_\key(H(\obj)^B\| \textrm{"key"}), C_{\sigma, j, i})) = 1$ but $\obj$ is not in the curator’s database.  In particular, this signature verifies only if $H(\obj) = H(\obj')$ for some $\obj'$ in the database, which means the client found a collision in the hash function.

\paragraph{\prooforigin} This follows from the unforgeability of the signature scheme, since clients confirm that $H(\obj)$ is signed by the curators.  In particular, a client that is willing to reveal its possession of an object $\obj$ can provide public evidence that it was under enforcement by revealing $\obj$ and $\{\sigma_j\}_j$.

\paragraph{\proofconsistency} Intuitively, this follows from the fact that the hash of the database is stored in an append-only log with an authoritative root.  In more depth, the gossip protocol used for the checkpoints of the Merkle tree $T$ guarantees that the checkpoints seen by all clients are either the same or consistent with each other, in terms of representing a single append-only log.  
The only way in which two clients interacting with the enforcer at the same point in time could come to a different verdict on the same object is thus if (1) the enforcer uses an outdated dataset for one client but a new one for the other or (2) the enforcer uses the latest dataset for the first client but then adds a new dataset to the log before interacting with the second client.  Assuming collision resistance, the first option is not possible due to the client's check of the Merkle inclusion proof; i.e., the client cannot be falsely convinced that its database is the rightmost entry in the tree.  The second option is not prevented, but can later be detected by even an unprivileged auditor, who can see that the enforcer has violated its policy in terms of updating its dataset multiple times within the same interval.  Assuming efficient reporting channels and sufficient disincentives for this type of behavior (e.g., having all users leave the platform), we can thus conclude that this option will not happen either.

\paragraph{\proofaudit} The leaf hash in the Merkle tree serves as a binding commitment to the contents of the database, which means it is computationally infeasible (by pre-image resistance) for any party to find a different set of blocklist elements that is represented by the same commitment.  Furthermore, by the same argument as above the view the client has of the Merkle tree is authoritative and globally consistent --- any commitment the client has seen will also be seen by an auditor. 

\paragraph{\realtime} Per our benchmarks in Section~\ref{sec:evaluation}, our time-efficient protocol costs only microseconds of computation on a client, with all other costs dominated by network latency associated with a single round-trip request between the client and enforcer for the blinding operation. Conversely, our space-efficient protocol needs to download 92.9KB (for $N=4096$), along with expending 1.4 seconds of computation on the server for a blocklist of roughly 873,000 elements (and significantly more for even larger blocklists). While parallelization of this server-side computation is possible, two round-trip requests and the associated download size may not be practical for many applications, hence why we avoid asserting that this protocol is \emph{\realtime} at this time.

\paragraph{\resistdos} Our time-efficient protocol uses nearly identical computational resources for the client and server as shown in Section~\ref{sec:evaluation}---even for malformed requests---and thus satisfies this property. Conversely, our space-efficient protocol uses orders of magnitude more computation on the part of the server compared to the client. Furthermore, a malicious client could avoid the bulk of the computation costs by sending bogus PIR indexes to the server. As such, we avoid saying that the latter has \emph{\resistdos} at this time.

%% file: main.bbl
\begin{thebibliography}{10}

\bibitem{aes_gcm_rust}
{AEADs}/aes-gcm.
\newblock \url{https://github.com/RustCrypto/AEADs}, 2022.

\bibitem{google_benchmark}
Benchmark.
\newblock \url{https://github.com/google/benchmark}, June 2022.

\bibitem{dalek_curve25519}
curve25519-dalek.
\newblock \url{https://github.com/dalek-cryptography/curve25519-dalek}, 2022.

\bibitem{dalek_ed25519}
ed25519-dalek.
\newblock \url{https://github.com/dalek-cryptography/ed25519-dalek}, 2022.

\bibitem{FTC_ZOOM}
{FTC} {Requires} {Zoom} to {Enhance} its {Security} {Practices} as {Part} of
  {Settlement}.
\newblock
  \url{http://www.ftc.gov/news-events/news/press-releases/2020/11/ftc-requires-zoom-enhance-its-security-practices-part-settlement},
  2022.

\bibitem{SealPIR}
{SealPIR}: {A} computational {PIR} library that achieves low communication
  costs and high performance.
\newblock \url{https://github.com/microsoft/SealPIR}, April 2022.

\bibitem{abelson_bugs_2021}
Hal Abelson, Ross Anderson, Steven~M. Bellovin, Josh Benaloh, Matt Blaze, Jon
  Callas, Whitfield Diffie, Susan Landau, Peter~G. Neumann, Ronald~L. Rivest,
  Jeffrey~I. Schiller, Bruce Schneier, Vanessa Teague, and Carmela Troncoso.
\newblock Bugs in our {Pockets}: {The} {Risks} of {Client}-{Side} {Scanning}.
\newblock {\em arXiv:2110.07450 [cs]}, 2021.

\bibitem{kud}
Harold Abelson, Ross Anderson, Steven~M Bellovin, Josh Benaloh, Matt Blaze,
  Whitfield Diffie, John Gilmore, Matthew Green, Susan Landau, Peter~G Neumann,
  et~al.
\newblock Keys under doormats: mandating insecurity by requiring government
  access to all data and communications.
\newblock {\em Journal of Cybersecurity}, 2015.

\bibitem{akhawe2013alice}
Devdatta Akhawe and Adrienne~Porter Felt.
\newblock Alice in warningland: A large-scale field study of browser security
  warning effectiveness.
\newblock In {\em Proceedings of the USENIX Security Symposium}, 2013.

\bibitem{contour}
Mustafa Al-Bassam and Sarah Meiklejohn.
\newblock Contour: A practical system for binary transparency.
\newblock In {\em Proceedings of the 2nd International Workshop on
  Cryptocurrencies and Blockchain Technology (CBT)}, 2018.

\bibitem{alhaddad_hecate_2021}
Rawane~Issa Alhaddad, Nicolas and Mayank Varia.
\newblock Hecate: {Abuse} {Reporting} in {Secure} {Messengers} with {Sealed}
  {Sender}, 2021.
\newblock Report Number: 1686.

\bibitem{mulpir}
Asra Ali, Tancr{\`e}de Lepoint, Sarvar Patel, Mariana Raykova, Phillipp
  Schoppmann, Karn Seth, and Kevin Yeo.
\newblock {Communication{\textendash}Computation} trade-offs in {PIR}.
\newblock In {\em Proceedings of the USENIX Security Symposium}, 2021.

\bibitem{messenger-encryption}
Ron Amadeo.
\newblock {Google} enables end-to-end encryption for {Android’s} default
  sms/rcs app.
\newblock
  \url{https://arstechnica.com/gadgets/2021/06/google-enables-end-to-end-encryption-for-androids-default-sms-rcs-app/},
  2021.

\bibitem{appleneuralhash}
Apple.
\newblock {CSAM} detection.
\newblock
  \url{https://www.apple.com/child-safety/pdf/CSAM_Detection_Technical_Summary.pdf},
  2021.

\bibitem{reverse-photodna}
Anish Athalye.
\newblock Inverting {PhotoDNA}.
\newblock \url{https://www.anishathalye.com/2021/12/20/inverting-photodna/},
  2021.

\bibitem{aumann07tcc}
Yonatan Aumann and Yehuda Lindell.
\newblock Security against covert adversaries: Efficient protocols for
  realistic adversaries.
\newblock In {\em Proceedings of the Theory of Cryptography Conference (TCC)},
  pages 137--156, 2007.

\bibitem{persona}
Randy Baden, Adam Bender, Neil Spring, Bobby Bhattacharjee, and Daniel Starin.
\newblock Persona: An online social network with user-defined privacy.
\newblock In {\em Proceedings of the ACM SIGCOMM Conference on Data
  Communication}, 2009.

\bibitem{bai2020targeted}
Jiawang Bai, Bin Chen, Yiming Li, Dongxian Wu, Weiwei Guo, Shu-tao Xia, and
  En-hui Yang.
\newblock Targeted attack for deep hashing based retrieval.
\newblock In {\em European Conference on Computer Vision}, 2020.

\bibitem{barenghi_snake_2014}
Alessandro Barenghi, Michele Beretta, Alessandro Di~Federico, and Gerardo
  Pelosi.
\newblock Snake: {An} {End}-to-{End} {Encrypted} {Online} {Social} {Network}.
\newblock In {\em {IEEE} {Intl} {Conf} on {High} {Performance} {Computing} and
  {Communications}}, 2014.

\bibitem{MLS}
Richard Barnes, Benjamin Beurdouche, Raphael Robert, Jon Millican, Emad Omara,
  and Katriel Cohn-Gordon.
\newblock The {Messaging} {Layer} {Security} ({MLS}) {Protocol}.
\newblock \url{https://datatracker.ietf.org/doc/draft-ietf-mls-protocol}, 2022.

\bibitem{bhowmick_apple_nodate}
Abhishek Bhowmick, Dan Boneh, Steve Myers, Kunal Talwar, and Karl Tarbe.
\newblock The {Apple} {PSI} {System}.
\newblock
  \url{https://www.apple.com/child-safety/pdf/Apple_PSI_System_Security_Protocol_and_Analysis.pdf},
  2021.

\bibitem{boneh2001short}
Dan Boneh, Ben Lynn, and Hovav Shacham.
\newblock Short signatures from the weil pairing.
\newblock In {\em International conference on the theory and application of
  cryptology and information security}. Springer, 2001.

\bibitem{android-caller-id}
Christina Bonnington.
\newblock How {Android} is fighting spam calls.
\newblock
  \url{https://slate.com/technology/2018/07/spam-calls-how-google-is-fighting-robocalls-on-android.html},
  2018.

\bibitem{bursztein2019rethinking}
Elie Bursztein, Einat Clarke, Michelle DeLaune, David~M Elifff, Nick Hsu,
  Lindsey Olson, John Shehan, Madhukar Thakur, Kurt Thomas, and Travis Bright.
\newblock Rethinking the detection of child sexual abuse imagery on the
  internet.
\newblock In {\em Proceedings of The Web Conference}, 2019.

\bibitem{chor1995private}
Benny Chor, Oded Goldreich, Eyal Kushilevitz, and Madhu Sudan.
\newblock Private information retrieval.
\newblock In {\em Proceedings of the Annual Symposium on Foundations of
  Computer Science}, 1995.

\bibitem{giftc-database}
Elizabeth Culliford.
\newblock {Facebook} and tech giants to target attacker manifestos, far-right
  militias in database.
\newblock
  \url{https://www.reuters.com/technology/exclusive-facebook-tech-giants-target-manifestos-militias-database-2021-07-26/},
  2021.

\bibitem{dolhansky2020adversarial}
Brian Dolhansky and Cristian~Canton Ferrer.
\newblock Adversarial collision attacks on image hashing functions.
\newblock {\em arXiv preprint arXiv:2011.09473}, 2020.

\bibitem{CCS:FDPFSS14}
Sascha Fahl, Sergej Dechand, Henning Perl, Felix Fischer, Jaromir Smrcek, and
  Matthew Smith.
\newblock Hey, {NSA}: Stay away from my market! future proofing app markets
  against powerful attackers.
\newblock In {\em Proceedings of ACM CCS}, 2014.

\bibitem{feldman_social_2012}
Ariel~J. Feldman, Aaron Blankstein, Michael~J. Freedman, and Edward~W. Felten.
\newblock Social {Networking} with {Frientegrity}: {Privacy} and {Integrity}
  with an {Untrusted} {Provider}.
\newblock In {\em Proceedings of the USENIX Security Symposium}, 2012.

\bibitem{goltzsche_endbox_2018}
David Goltzsche, Signe Rusch, Manuel Nieke, Sebastien Vaucher, Nico Weichbrodt,
  Valerio Schiavoni, Pierre-Louis Aublin, Paolo Cosa, Christof Fetzer, Pascal
  Felber, Peter Pietzuch, and Rudiger Kapitza.
\newblock {EndBox}: {Scalable} {Middlebox} {Functions} {Using} {Client}-{Side}
  {Trusted} {Execution}.
\newblock In {\em {IEEE}/{IFIP} {International} {Conference} on {Dependable}
  {Systems} and {Networks} ({DSN})}, 2018.

\bibitem{google_ai_principles}
Google.
\newblock Our principles.

\bibitem{haitner2008linear}
Iftach Haitner, Jonathan~J Hoch, and Gil Segev.
\newblock A linear lower bound on the communication complexity of single-server
  private information retrieval.
\newblock In {\em Proceedings of the Theory of Cryptography Conference}, 2008.

\bibitem{hale_komlo}
Britta Hale and Chelsea Komlo.
\newblock On {End}-to-{End} {Encryption}.
\newblock \url{https://eprint.iacr.org/2022/449}, 2022.

\bibitem{han2017sgx}
Juhyeng Han, Seongmin Kim, Jaehyeong Ha, and Dongsu Han.
\newblock Sgx-box: Enabling visibility on encrypted traffic using a secure
  middlebox module.
\newblock In {\em Proceedings of the First Asia-Pacific Workshop on
  Networking}, pages 99--105, 2017.

\bibitem{hao2021s}
Qingying Hao, Licheng Luo, Steve~TK Jan, and Gang Wang.
\newblock It's not what it looks like: Manipulating perceptual hashing based
  applications.
\newblock In {\em Proceedings of the 2021 ACM SIGSAC Conference on Computer and
  Communications Security}, 2021.

\bibitem{criterion}
Brook Heisler.
\newblock Criterion.rs.
\newblock \url{https://github.com/bheisler/criterion.rs}, June 2022.

\bibitem{ncmec-database}
Alex Hern.
\newblock Sites reported record 29.3m child abuse images in 2021.
\newblock
  \url{https://www.theguardian.com/technology/2022/mar/24/sites-reported-record-293m-child-abuse-images-in-2021},
  2022.

\bibitem{hitchen_analysis_2019}
Jamie Hitchen, Jonathan Fisher, Nic Cheeseman, and Idayat Hassan.
\newblock Analysis {\textbar} {How} {WhatsApp} influenced {Nigeria}’s recent
  election — and what it taught us about ‘fake news.’.
\newblock {\em Washington Post}, February 2019.

\bibitem{horel_how_2018}
Thibaut Horel, Sunoo Park, Silas Richelson, and Vinod Vaikuntanathan.
\newblock How to {Subvert} {Backdoored} {Encryption}: {Security} {Against}
  {Adversaries} that {Decrypt} {All} {Ciphertexts}.
\newblock page 20 pages, 2018.
\newblock arXiv:1802.07381 [cs].

\bibitem{hua2021increasing}
Yiqing Hua, Armin Namavari, Kaishuo Cheng, Mor Naaman, Thomas Ristenpart, and
  Cornell Tech.
\newblock Increasing adversarial uncertainty to scale private similarity
  testing.
\newblock {\em Proceedings of the USENIX Security Symposium}, 2022.

\bibitem{huberman1999enhancing}
Bernardo~A Huberman, Matt Franklin, and Tad Hogg.
\newblock Enhancing privacy and trust in electronic communities.
\newblock In {\em Proceedings of the ACM Conference on Electronic Commerce},
  1999.

\bibitem{jain2021adversarial}
Shubham Jain, Ana-Maria Cretu, and Yves-Alexandre de~Montjoye.
\newblock Adversarial detection avoidance attacks: Evaluating the robustness of
  perceptual hashing-based client-side scanning.
\newblock In {\em NeurIPS 2021 Workshop Privacy in Machine Learning}, 2021.

\bibitem{kamara_outside_2022}
Seny Kamara, Mallory Knodel, Emma Llansó, Greg Nojeim, Lucy Qin, Dhanaraj
  Thakur, and Caitlin Vogus.
\newblock Outside {Looking} {In}: {Approaches} to {Content} {Moderation} in
  {End}-to-{End} {Encrypted} {Systems}.
\newblock {\em arXiv:2202.04617 [cs]}, 2022.

\bibitem{whatsapp-forwarding}
Jacob Kastrenakes.
\newblock Whatsapp limits message forwarding in fight against misinformation.
\newblock
  \url{https://www.theverge.com/2019/1/21/18191455/whatsapp-forwarding-limit-five-messages-misinformation-battle},
  2019.

\bibitem{knodel}
Mallory Knodel, Fred Baker, Olaf Kolkman, Sofia Celi, and Gurshabad Grover.
\newblock Definition of {End}-to-end {Encryption}.
\newblock \url{https://datatracker.ietf.org/doc/draft-knodel-e2ee-definition},
  2022.

\bibitem{kogan2021private}
Dmitry Kogan and Henry Corrigan-Gibbs.
\newblock Private blocklist lookups with checklist.
\newblock In {\em Proceedings of the USENIX Security Symposium}, 2021.

\bibitem{kulshrestha2021identifying}
Anunay Kulshrestha and Jonathan Mayer.
\newblock Identifying harmful media in {End-to-End} encrypted communication:
  Efficient private membership computation.
\newblock In {\em Proceedings of the USENIX Security Symposium}, 2021.

\bibitem{kwon_riffle_2016}
Albert Kwon, David Lazar, Srinivas Devadas, and Bryan Ford.
\newblock Riffle: {An} {Efficient} {Communication} {System} {With} {Strong}
  {Anonymity}.
\newblock {\em Proceedings on Privacy Enhancing Technologies}, 2016.

\bibitem{photodna}
Jennifer Langston.
\newblock How {PhotoDNA} for video is being used to fight online child
  exploitation.
\newblock
  \url{https://news.microsoft.com/on-the-issues/2018/09/12/how-photodna-for-video-is-being-used-to-fight-online-child-exploitation/},
  2018.

\bibitem{larisch2017crlite}
James Larisch, David Choffnes, Dave Levin, Bruce~M Maggs, Alan Mislove, and
  Christo Wilson.
\newblock Crlite: A scalable system for pushing all tls revocations to all
  browsers.
\newblock In {\em IEEE Symposium on Security and Privacy (SP)}, 2017.

\bibitem{lazar_alpenhorn_2016}
David Lazar and Nickolai Zeldovich.
\newblock Alpenhorn: Bootstrapping secure communication without leaking
  metadata.
\newblock In {\em Proceedings of the USENIX Symposium on Operating Systems
  Design and Implementation}, 2016.

\bibitem{robinsonlevy}
Ian Levy and Crispin Robinson.
\newblock Thoughts on child safety on commodity platforms, July 2022.
\newblock arXiv:2207.09506 [cs].

\bibitem{li2019protocols}
Lucy Li, Bijeeta Pal, Junade Ali, Nick Sullivan, Rahul Chatterjee, and Thomas
  Ristenpart.
\newblock Protocols for checking compromised credentials.
\newblock \url{https://rist.tech.cornell.edu/papers/c3.pdf}, 2019.

\bibitem{liu2021fighting}
Linsheng Liu, Daniel~S Roche, Austin Theriault, and Arkady Yerukhimovich.
\newblock Fighting fake news in encrypted messaging with the fuzzy anonymous
  complaint tally system (facts).
\newblock {\em arXiv preprint arXiv:2109.04559}, 2021.

\bibitem{mayrhofer2021android}
Ren{\'e} Mayrhofer, Jeffrey~Vander Stoep, Chad Brubaker, and Nick Kralevich.
\newblock The android platform security model.
\newblock {\em ACM Transactions on Privacy and Security (TOPS)}, 24(3):1--35,
  2021.

\bibitem{meiklejohn2020gossip}
Sarah Meiklejohn, Pavel Kalinnikov, Cindy~S Lin, Martin Hutchinson, Gary
  Belvin, Mariana Raykova, and Al~Cutter.
\newblock Think global, act local: Gossip and client audits in verifiable data
  structures.
\newblock {\em arXiv preprint arXiv:2011.04551}, 2020.

\bibitem{muffett}
Alec Muffett.
\newblock A {Duck} {Test} for {End}-to-{End} {Secure} {Messaging}.
\newblock
  \url{https://datatracker.ietf.org/doc/draft-muffett-end-to-end-secure-messaging-02},
  2022.

\bibitem{wired-gbt}
Lily~Hay Newman.
\newblock The {Pixel} 6 {Tensor} chip's best upgrade isn't speed. it's
  security, October 2021.
\newblock
  \url{https://www.wired.com/story/google-pixel-6-tensor-chip-security/}.

\bibitem{facebook-pdq}
Casey Newton.
\newblock {Facebook} open-sources algorithms for detecting child exploitation
  and terrorism imagery.
\newblock
  \url{https://www.theverge.com/2019/8/1/20750752/facebook-child-exploitation-terrorism-open-source-algorithm-pdq-tmk},
  2019.

\bibitem{chainiac}
Kirill Nikitin, Eleftherios Kokoris-Kogias, Philipp Jovanovic, Nicolas Gailly,
  Linus Gasser, Ismail Khoffi, Justin Cappos, and Bryan Ford.
\newblock {CHAINIAC}: Proactive software-update transparency via collectively
  signed skipchains and verified builds.
\newblock In {\em Proceedings of the 26th USENIX Security Symposium}, 2017.

\bibitem{signal}
Jim Oleary.
\newblock Improving first impressions on {Signal}.
\newblock \url{https://signal.org/blog/keeping-spam-off-signal/}, 2021.

\bibitem{pereira2020metadata}
Mayana Pereira, Rahul Dodhia, Hyrum Anderson, and Richard Brown.
\newblock Metadata-based detection of child sexual abuse material.
\newblock {\em arXiv preprint arXiv:2010.02387}, 2020.

\bibitem{chrome-malware-warning}
Jay Peters.
\newblock {Google’s Advanced Protection Program} will protect against risky
  {Chrome} downloads.
\newblock
  \url{https://www.theverge.com/2019/8/6/20757081/google-advanced-protection-program-risky-chrome-downloads-protection},
  2019.

\bibitem{pfefferkorn_content-oblivious_2021}
Riana Pfefferkorn.
\newblock Content-oblivious trust and safety techniques: Results from a survey
  of online service providers.

\bibitem{pinkas2014faster}
Benny Pinkas, Thomas Schneider, and Michael Zohner.
\newblock Faster private set intersection based on ot extension.
\newblock In {\em Proceedings of the USENIX Security Symposium}, 2014.

\bibitem{ramachandran2007filtering}
Anirudh Ramachandran, Nick Feamster, and Santosh Vempala.
\newblock Filtering spam with behavioral blacklisting.
\newblock In {\em Proceedings of the ACM conference on computer and
  communications security}, 2007.

\bibitem{riazi2016sub}
M~Sadegh Riazi, Beidi Chen, Anshumali Shrivastava, Dan Wallach, and Farinaz
  Koushanfar.
\newblock Sub-linear privacy-preserving near-neighbor search.
\newblock {\em arXiv preprint arXiv:1612.01835}, 2016.

\bibitem{clark}
J.~H. Saltzer, D.~P. Reed, and D.~D. Clark.
\newblock End-to-end arguments in system design.
\newblock {\em ACM Transactions on Computer Systems}, 1984.

\bibitem{sherry_blindbox_2015}
Justine Sherry, Chang Lan, Raluca~Ada Popa, and Sylvia Ratnasamy.
\newblock {BlindBox}: {Deep} {Packet} {Inspection} over {Encrypted} {Traffic}.
\newblock In {\em Proceedings of the 2015 {ACM} {Conference} on {Special}
  {Interest} {Group} on {Data} {Communication}}, London United Kingdom, 2015.

\bibitem{singh2019robust}
Priyanka Singh and Hany Farid.
\newblock Robust homomorphic image hashing.
\newblock In {\em CVPR Workshops}, 2019.

\bibitem{struppek2021learning}
Lukas Struppek, Dominik Hintersdorf, Daniel Neider, and Kristian Kersting.
\newblock Learning to break deep perceptual hashing: The use case neuralhash.
\newblock {\em arXiv preprint arXiv:2111.06628}, 2021.

\bibitem{syta2016keeping}
Ewa Syta, Iulia Tamas, Dylan Visher, David~Isaac Wolinsky, Philipp Jovanovic,
  Linus Gasser, Nicolas Gailly, Ismail Khoffi, and Bryan Ford.
\newblock Keeping authorities" honest or bust" with decentralized witness
  cosigning.
\newblock In {\em Proceedings of the IEEE Symposium on Security and Privacy},
  2016.

\bibitem{thomas2019protecting}
Kurt Thomas, Jennifer Pullman, Kevin Yeo, Ananth Raghunathan, Patrick~Gage
  Kelley, Luca Invernizzi, Borbala Benko, Tadek Pietraszek, Sarvar Patel, Dan
  Boneh, et~al.
\newblock Protecting accounts from credential stuffing with password breach
  alerting.
\newblock In {\em Proceedings of the USENIX Security Symposium}, 2019.

\bibitem{topalovic2012towards}
Emin Topalovic, Brennan Saeta, Lin-Shung Huang, Collin Jackson, and Dan Boneh.
\newblock Towards short-lived certificates.
\newblock In {\em Proceedings of the Workshop on Web 2.0 Security and Privacy},
  2012.

\bibitem{tyagi2019asymmetric}
Nirvan Tyagi, Paul Grubbs, Julia Len, Ian Miers, and Thomas Ristenpart.
\newblock Asymmetric message franking: content moderation for metadata-private
  end-to-end encryption.
\newblock In {\em Annual International Cryptology Conference}. Springer, 2019.

\bibitem{tyagi2019traceback}
Nirvan Tyagi, Ian Miers, and Thomas Ristenpart.
\newblock Traceback for end-to-end encrypted messaging.
\newblock In {\em Proceedings of the Conference on Computer and Communications
  Security}, 2019.

\bibitem{van2018foreshadow}
Jo~Van~Bulck, Marina Minkin, Ofir Weisse, Daniel Genkin, Baris Kasikci, Frank
  Piessens, Mark Silberstein, Thomas~F Wenisch, Yuval Yarom, and Raoul Strackx.
\newblock Foreshadow: Extracting the keys to the intel {SGX} kingdom with
  transient out-of-order execution.
\newblock In {\em Proceedings of the {USENIX} Security Symposium}, 2018.

\bibitem{wang_cloak_2011}
David~Y. Wang, Stefan Savage, and Geoffrey~M. Voelker.
\newblock Cloak and dagger: dynamics of web search cloaking.
\newblock In {\em Proceedings of the {ACM} conference on {Computer} and
  communications security}, 2011.

\bibitem{wang2021prototype}
Xunguang Wang, Zheng Zhang, Baoyuan Wu, Fumin Shen, and Guangming Lu.
\newblock Prototype-supervised adversarial network for targeted attack of deep
  hashing.
\newblock In {\em Proceedings of the IEEE/CVF Conference on Computer Vision and
  Pattern Recognition}, 2021.

\bibitem{apple_icloud_e2ee}
Kyle Wiggers.
\newblock Apple launches end-to-end encryption for {iCloud} data.
\newblock
  \url{https://techcrunch.com/2022/12/07/apple-launches-end-to-end-encryption-for-icloud-data/},
  2022.

\bibitem{facebook-encrypted}
Mark Zuckerberg.
\newblock A privacy-focused vision for social networking.
\newblock \url{https://www.facebook.com/notes/2420600258234172/}, 2021.

\end{thebibliography}
